\documentclass[aps,a4paper, preprint, superscriptaddress,preprintnumbers,
nofootinbib,amsmath,amssymb]{revtex4}
\usepackage{epsfig}
\usepackage{amsfonts}
\usepackage{url}
\usepackage{subfigure}
\usepackage{hhline}
\usepackage{bm}
\usepackage{comment}
\usepackage[usenames]{color}
\usepackage{minitoc}
\usepackage{tabulary}
\usepackage{array}
\usepackage{enumerate}
\usepackage{acronym}
\usepackage{pdflscape}
\usepackage{multirow}
\usepackage{doi}
\usepackage{hyperref}
\usepackage{graphicx}
\hypersetup{colorlinks=true, linkcolor=blue,citecolor=blue,urlcolor=blue}
\begin{document}
\title{$A_{5}$ symmetry and deviation from Golden Ratio mixing with charged lepton flavor violation}

\author{Victoria Puyam}
\email{victoria.phd.phy@manipuruniv.ac.in}
 \affiliation{Department of Physics, Manipur University, Imphal-795003, India}                                      
  
  \author{ N. Nimai Singh}
\email{nimai03@yahoo.com}
 \affiliation{Department of Physics, Manipur University, Imphal-795003, India}   
   \affiliation{  Research Institute of Science and Technology, Imphal-795003, India }  

\begin{abstract}
 A neutrino mass model that can satisfy the exact golden ratio mixing is constructed using $A_{5}$ discrete symmetry group. The deviation from the golden ratio mixing is studied by considering the contribution from the charged lepton sector in a linear seesaw framework. A definite pattern of charged lepton mass matrix predicted by the model controls the leptonic mixing angles. By taking the observed $\theta_{13}$ as the input value, we can obtain the values of all the mixing angles and Dirac CP-violating phase within the current experimental bounds.

The model predicts that only the normal neutrino mass ordering is consistent with the current oscillation data. We also study the two body charged lepton flavor violation (cLFV) processes such as  $\mu \rightarrow e +\gamma$, $\tau \rightarrow e +\gamma$  and $\tau \rightarrow \mu+ \gamma$ and neutrinoless double beta decay parameter $m_{\beta \beta}$. The present neutrino mass model can explain the current and future sensitivity of $\mu \rightarrow e +\gamma$, $\tau \rightarrow e+ \gamma$ processes and the present sensitivity of neutrinoless double beta decay parameter when the masses of quasi-Dirac neutrinos are in the TeV range. On the other hand, the model cannot reproduce the present sensitivity of $\tau \rightarrow \mu+ \gamma$  but can explain the future sensitivity and the present sensitivity of neutrinoless double beta decay parameter simultaneously when the masses of the quasi-Dirac neutrinos are in the TeV range.

\end{abstract}
\maketitle
\ifpdf
    \graphicspath{{ch 1/Fig/Raster/}{ch 1/Fig/EPS/}{ch 1/Figs/}}
\else
    \graphicspath{{ch 1/Fig/Vector/}{ch 1/Fig/}}
\fi
\section{Introduction}
Present knowledge on neutrino physics comes from neutrino oscillation experiments \cite{KamLAND:2002uet,SNO:2002tuh,Super-Kamiokande:1998kpq,DoubleChooz:2011ymz,Lasserre:2012ax,Ling:2013fta,McDonald:2016ixn}, neutrinoless double beta decay experiments \cite{GERDA:2019ivs,CUORE:2019yfd,KamLAND-Zen:2016pfg}, the end point of tritium beta decay experiment \cite{KATRIN:2019yun} and cosmological observations \cite{Planck:2018nkj}. The precision measurement of the leptonic mixing angles in the Pontecorvo-Maki-Nakagawa-Sakata (PMNS) matrix $U_{\text{PMNS}}$, shows that $\theta_{23}$ and $\theta_{12}$ are large and $\theta_{13}$ is small but non-zero. The upper bound on the sum of the three neutrino masses $\sum m_{i}\leq 0.12$ eV, is also observed from the Planck cosmological observations \cite{Planck:2018nkj}. The current global fit on neutrino oscillation data is given in Table \ref{t3}. 
\begin{table}[h]
  \begin{center}
\begin{tabular}{cccc}
\hline
\hline 
 Parameters& Best fit$\pm $1$\sigma$& 2$\sigma$ &3$\sigma$ \\
\hline
\hline
\vspace*{0.2 cm}
$\theta_{12}/^{\circ}$ &$ 34.3\pm 1.0$& 32.3-36.4&31.4-37.4   \\  
\vspace*{0.2 cm}
$\theta_{13}/^{\circ}$(NO) &$8.53^{+0.13}_{-0.12}$&8.27-8.79& 8.20-8.97   \\ 
\vspace*{0.2 cm}
$\theta_{13}/^{\circ}$(IO) &$8.58^{+0.12}_{-0.14}$&8.30-8.83 &8.17-8.96   \\ 
\vspace*{0.2 cm}
$\theta_{23}/^{\circ}$(NO) &$49.26\pm 0.79$&47.37-50.71& 41.20-51.33   \\ 
\vspace*{0.2 cm}
$\theta_{23}/^{\circ}$(IO) &$49.46^{+ 0.60}_{-0.97}$&47.35-50.67& 41.16-51.25   \\ 
\vspace*{0.2 cm}
 $\Delta m_{21}^{2}[10^{-5}eV^{2}]$& $7.50^{+ 0.22}_{-0.20} $ &7.12-7.93& 6.94-8.14\\ 
\vspace*{0.2 cm} 
$|\Delta m_{31}^{2}|[10^{-3}eV^{2}]$(NO) &$2.55^{+0.02}_{-0.03}$& 2.49-2.60&2.47-2.63 \\
\vspace*{0.2 cm} 
 $|\Delta m_{31}^{2}|[10^{-3}eV^{2}]$(IO) &$2.45^{+0.02}_{-0.03}$&2.39-2.50&2.37-2.53 \\
\vspace{0.2 cm}
$\delta/^{\circ}$(NO)& $194^{\footnotesize{+24}}_{\footnotesize{-22}}$&152-255 & 128-359\\
\vspace*{0.2 cm}
$\delta/^{\circ}$(IO)& $284^{\footnotesize{+26}}_{\footnotesize{-28}}$ & 226-332&200-353\\
\hline
\hline
\end{tabular}
\caption{\footnotesize{The global-fit results for neutrino oscillation parameters \cite{deSalas:2020pgw}.}}
\label{t3}
 \end{center}
\end{table} 
 
 Such neutrino experiments data clearly hints at physics beyond the Standard Model (BSM). However, they still have shortcomings like octant of $\theta_{23}$, neutrino mass ordering, CP violation in the lepton sector, baryon asymmetry of the universe, etc. So, many new theories and experiments are pursued to study BSM. One such exciting approach is the charged lepton flavor violation (cLFV). These processes are highly suppressed in the Standard Model(SM). In the neutrino mass model based on the type-I seesaw framework,  expected magnitude of the cLFV processes are negligible due to the heavy masses of right-handed neutrino introduced to account for the smallness of active neutrino masses. However, sizeable structures of electroweak currents that contribute to the cLFV process, can be explained in a low-scale seesaw-like linear or inverse seesaw. There are several presently available and planned experiments for two cLFV decays ($l_{i}\rightarrow l_{j}\gamma)$ \cite{ch11ch11}, and three-body decays $(l_{i}\rightarrow l_{j}l_{k}l_{k})$ \cite{annurev-nucl-102912-144530}. In this work, we study the  cLFV decay $l_{i}\rightarrow l_{j}\gamma$ with a neutrino mass model based on a linear seesaw in the basis where the charged lepton mass matrix is non-diagonal.

On the other hand, the lepton masses and mixing patterns are often investigated by using several discrete symmetry groups such as $A_{4}$, $A_{5}$, $S_{3}$, $S_{4}$, $\Delta(96)$ etc \cite{Ding:2011cm,Feruglio:2011qq,Everett:2008et,CarcamoHernandez:2022bka,Lei:2020nik,BORAH201959,Puyam:2022mej,Puyam:2022efv,Gupta:2011ct,Gautam:2020bnx,Vien:2015fhk,CarcamoHernandez:2019eme,Kobayashi:2003fh,Morisi:2011pm,Feruglio:2019ybq,Ding:2023htn,Ding:2024ozt}. Out of these groups, the $A_{5}$ symmetry group is used to accommodate golden ratio(GR) mixing in a neutrino mass matrix. Here, it is interesting to note that in the GR mixing scenario, $\theta_{23}$ is maximal and $\theta_{13}=0$, while the solar mixing angle, $\theta_{12}$ is governed by the golden ratio, $\phi_{g}=(1+\sqrt{5})/2$ as $\tan \theta_{12}=1/\phi_{g}$ with $\sin^{2} \theta_{12}\approx 0.276$ in GR1 mixing or $\cos \theta_{12}=\phi_{g}/2$ with $\sin^{2} \theta_{12}\approx 0.345$ in GR2 mixing \cite{RODEJOHANN2009267,Ding:2011cm}. Both the GR mixing patterns are excluded by current data. As a result, the neutrino mass models based on GR mixing, need modifications. These deviations are obtained either by adding the contribution from the charged lepton sector \cite{CarcamoHernandez:2022bka,Cooper:2012bd,deMedeirosVarzielas:2013tbq,Gehrlein:2014wda} or by considering the vacuum alignment correction of the $A_{5}$ triplets or $A_{5}$ quintuplet \cite{Feruglio:2011qq,Ding:2011cm} or by including the extra contribution in the neutrino sector \cite{Cooper:2012bd,di2015lepton,PhysRevD.92.093008,PhysRevD.92.116007,li2015lepton,lopez2019lepton}.

In this work, we aim to construct a lepton mass model based on $A_{5} \times Z_{5} \times Z_{9}\times Z_{5}$ symmetry group that can produce the necessary deviation from GR1 mixing to accommodate the current neutrino masses and mixing data. Here, the light neutrino masses are generated through the linear seesaw mechanism, and its mass matrix retains the GR1 mixing pattern. On the other hand, a specific pattern of charged lepton mass matrix can perturb GR1 mixing and control the neutrino mixing pattern. It is interesting to note that there are earlier models with different structures of charged lepton mass matrix \cite{CarcamoHernandez:2022bka,deMedeirosVarzielas:2013tbq,Gehrlein:2014wda}. In the present work, we employ charged lepton correction in 12 sector, which is quite different from the charged lepton correction adopted in other works mentioned above. The way the charged lepton mass matrix is structured, has a significant impact on the lepton mixing matrix \cite{CarcamoHernandez:2022bka,deMedeirosVarzielas:2013tbq}. Here, unlike the earlier models based on $A_{5}$ symmetry, our model predicts only the normal ordering of neutrino masses. Again, the specific charged lepton mass matrix structure of our model allows us to study a distinct lepton mixing matrix and its related phenomenology. One such interesting prediction of the model is that the value of $\theta_{23}$ is found in the lower octant region for normal hierarchy (NH). In addition, the model also shows a strong correlation between $\theta_{13}$ and other mixing angles and Dirac CP-violating phase $\delta_{CP}$, as they can be obtained in their current bounds by taking the experimental values of the  $\theta_{13}$ as the input value. In addition, we also discuss the implication of the cLFV ($l_{i} \rightarrow l_{j}\gamma)$ processes to our model. Furthermore, we also study the neutrinoless double beta decay parameter $m_{\beta \beta}$ and its impact on cLFV sensitivity and model parameters. Here, we try to find the allowed space of our neutrino model parameters that explain all the current neutrino mass and oscillation parameters along with the sensitivity of cLFV and neutrinoless double beta decay.

The paper is organized as follows: In Section II, we describe the model and its particle contents in the underlying symmetry group and lepton mixing pattern arising from the model. Section III deals with  diagonalization process for charged lepton mass matrix and neutrino mass matrix. Section IV and V give some details of cLFV processes and neutrinoless double beta decay, respectively. The results of our numerical analysis are discussed in Section VI. In Section VII, we give the conclusion and summary of the present work. A brief sketch on $A_{5}$ group is relegated to Appendix A. 

\section{Description of the Model}
The lepton flavor model considered in the present work is based on $A_{5}$ symmetry group. Here, the fermion fields consist of the Standard Model (SM) left-handed lepton doublet $l$, the right- handed lepton singlet $e_{R}$, $\mu_{R}$ and $\tau_{R}$ and two right- handed SM model singlet neutrino fields $N_{R}$ and $S_{R}$. The SM lepton doublet and the neutrino fields have the representation $3$ under $A_{5}$ while the SM right handed singlets transform as singlet under $A_{5}$. The two right handed neutrino fields contribute to the masses of light active neutrinos through the linear seesaw mechanism.  The $A_{5}$ symmetry allows the inclusion of the SM singlet but $A_{5}$ pentaplet scalar $\phi_{5}$ in the model and it is necessary  to accommodate the GR mixing in the neutrino mass model. In addition to $\phi_{5}$,  the scalars sector consists of  Higgs doublets $h_{u}$ and $h_{d}$ and SM singlet scalar  $\xi_{1}$, $\xi_{2}$, $\xi_{3}$, $\eta_{1}$, $\eta_{2}$, $\phi_{1}$, $\phi_{2}$ and $\phi_{3}$. The scalar fields $\phi_{1}$, $\phi_{2}$ and $\phi_{3}$ have $A_{5}$ representation 3 and they are necessary to generate the charged  lepton mass matrix. The $A_{5}$ singlet fields $\xi_{1}$ and $\xi_{2}$ are added to produce the charged lepton correction in 12 sector. The scalars  $\xi_{3}$ and $\eta_{1}$ are $A_{5}$ singlets and they contribute to the neutrino mass matrix along with $\phi_{5}$.  The resultant light neutrino mass matrix can accommodate the GR mixing pattern. It is to be noted that $\eta_{2}$ is the auxillary field added to implement the desired vacuum alingment and it does not interact with the charged lepton field and neutrino fields at the leading order \cite{Feruglio:2011qq}. Here, we follow the consideration as in Ref.\cite{Feruglio:2011qq} where no driving fields are introduced to determine the VEV alinment of the triplets and pentaplet.  Inclusion of the large number of flavons in this model can be prevented if the model is based on modular $A_{5}$ symmetry group \cite{Novichkov:2018nkm, Ding:2019xna, deMedeirosVarzielas:2022ihu,Behera:2022wco,Abbas:2024bbv}. However this paper will deal with the unique lepton mass matrices arising as a consequence of interaction between the leptons fields and the additional flavon fields under $A_{5}$ discrete symmetry group and other supplementary discrete symmetry group.

The $A_{5}$ symmetry group is  further supplemented by the $Z_{5}\times Z_{9}\times Z_{5}$ group to achieve the desired pattern of the lepton mass matrices. The first $Z_{5}$ group is introduced to distinguish the three $A_{5}$ triplets which in turn ensures the correct scalar triplet to interact with the correct sector. The $Z_{9}$ group is introduced to accommodate different representation to the two neutrino fields and  it allows $\phi_{5}$ to interact with only one neutrino field in the leading order. Furthermore, the inclusion of another $Z_{5}$ group  prevent the introduction of unwanted  Majorana mass terms arising from the self interaction of the neutrino fields. These Majorana mass terms are allowed in higher order such as $\bar{N^{c}}_{R}N_{R}\xi_{3}^{4}\xi_{2}^{2}\phi_{5}/\lambda^{6}$ and $ \bar{S^{c}}_{R}S_{R} \xi_{3}\eta_{2}\xi_{2}^{2}/\lambda^{3}$.  The total factor $A_{5}\times Z_{5}\times Z_{9} \times Z_{5}$ contributes to contruction of the scalar potential. So, the inclusion of these symmetries  restrict the  introduction of the unwanted terms in the charged lepton mass matrix and neutrino mass matrix, which are otherwise allowed by $A_{5}$ symmetry, and hence necessary to obtain the specific pattern of the charged lepton mass matrix and neutrino mass matrix. The model is considered in supersymmetric (SUSY) framework as it allows one to naturally obtain the desired VEV alignment of the $A_5$ scalar triplets and the $A_5$ scalar quintuplet used in the model \cite{CarcamoHernandez:2022bka,Feruglio:2011qq}. It is to be noted that SUSY breaking effects do not affect lepton masses and mixing angles \cite{Feruglio:2011qq}. The transformation properties of the fields used in our model under the symmetry group so far considered, are provided in Table \ref{tp}. Here, the $Z_{5}$, $Z_{9}$ and $Z_{5}$ charges are in additive notation.

\begin{table}[h]
\begin{tabular}{|c|c|c|c|c|c|c|c|c|c|c|c|c|c|c|c|c|}
\hline 
Fields&~$l$~ & ~$e_{R}$~ &~ $\mu_{R}$~ & ~$\tau_{R}$~ &~ $N_{R}$~ & ~$S_{R}$~ & ~$H_{u,d}$~ & ~$\phi_{1}$~ & ~$\phi_{2}$~ & ~$\phi_{3}$~ & ~$\phi_{5}$~& ~$\xi_{1}$~ &~$\xi_{2}$~ &~$\xi_{3}$~ & ~$\eta_{1}$~ & ~$\eta_{2}$~ \\ 
\hline 
$A_{5}$&3 & 1 &1 & 1 & ~3~ & ~3~ & ~1~ &~3~ & ~3~ & ~3~ & ~5~ & ~1~ & ~1~ & ~1~ & ~1~ & ~1~ \\ 
\hline 
 $Z_{5}$& 0 & -2 &~-1~ & ~1~ & ~0~ & ~0~ & ~0~ & ~0~ & ~1~ & ~4~ & ~0~ &~1~ & ~0~ & ~0~ & ~0~ & ~0~ \\ 
\hline 
$Z_{9}$&4 &-6 & ~-5~ & ~-4~ & ~-7~ & ~5~ & ~0~ & ~0~ & ~0~ & ~0~ & ~3~ & ~1~ & ~1~ & ~0~ & ~3~ & ~6~ \\ 
\hline 
$Z_{5}$&2&-2&-2&-2&3&2&0&0&0&0&0&0&0&1&0&0\\
\hline
\end{tabular} 
\caption{\footnotesize{Transformation properties of various fields under symmetry group considered in the model.}}
\label{tp}
\end{table}

The Yukawa Langrangian terms for the charged leptons and neutrino which are invariant under $A_{5}\times Z_{5} \times Z_{9}\times Z_{5}$ transformation are given below,
\begin{align}
-L_{y}=&y_{1}(\bar{l}~\frac{\phi_{1}}{\lambda^{3}}e_{R})h_{d}\xi^{2}_{1}+y_{2}(\bar{l}~\frac{\phi_{2}}{\lambda^{2}}\mu_{R})h_{d}\xi_{2}+y_{3}(\bar{l}~\frac{\phi_{3}}{\lambda}\tau_{R})h_{d}+y_{4}(\bar{l}~\frac{\phi_{2}}{\lambda^{3}}e_{R})h_{d}\xi_{1}\xi_{2}\nonumber\\ 
&+y_{5}(\bar{l}~\frac{\phi_{1}}{\lambda^{2}}\mu_{R})h_{d}\xi_{1} 
+\frac{y_{a}}{\lambda} (\bar{l}~N_{R} \phi_{5})_{1}h_{u}+\frac{y_{b}}{\lambda} (\bar{l}~N_{R} \eta_{1})_{1}h_{u}+\frac{y_{c}}{\lambda}(\bar{l}~ S_{R})_{1}h_{u}\xi_{3}+\nonumber\\
&\frac{y_{N}}{\lambda}(\bar{N}^{c}_{R}S_{R})\xi_{2}^{2}+h.c.,
\end{align}
where $\lambda$ is the model cut-off high scale. Here, the vacuum expectation value (VEV) pattern for the  scalars  used in the model are given in Table \ref{t2} . The VEV patterns for the $A_{5}$ scalar triplets and pentaplets considered in our model, can be obtained from the superpotential given below
\begin{align}
W=&M_{o}\eta_{1} \eta_{2}+ M_{1} \phi_{1}^{2}+M_{2} (\phi_{2}\phi_{3})+g(\phi_{1}\phi_{2}\phi_{3})+g_{1}\eta_{1}(\phi_{5}^{2})+g_{2}(\phi_{5}^{3})_{1}\nonumber\\ 
&+g_{3}(\phi_{5}^{3})_{2}+\frac{g_{4}}{3}\eta_{1}^3+\frac{g_{5}}{3}\eta^{3}_{2}+...
\label{w1}
\end{align}
 Here, terms like $((\phi_{1} \phi_{1})_{3}\phi_{1})_{1}$ are omitted in the superpotential since the contraction $(\phi_{1} \phi_{1})_{3}$ vanishes. The detailed calculations are shown in Appendix B. 
\begin{table}[h]
\begin{tabular}{cccc}

\hline
\hline
\vspace{0.1cm}
Scalar& && VEV \\
\hline
\hspace{0.1cm}
$<\phi_{1}>$&& & $v_{1}$ (1, 0, 0)\\
$<\phi_{2}>$&& &$v_{2}$ (0, 0, 1)\\
$<\phi_{3}>$&& &$v_{3}$ (0, 1, 0)\\
$<\phi_{5}>$&& &$v_{s}(-\sqrt{\frac{2}{3}}(p+q)$, $-p$, $q$, $q$, $p$) \\
$<h_{u,d}>$ &&& $v_{u,d}$\\
$<\eta_{1}>$ &&& $u$\\
$<\xi_{1}>$ &&& $u_{1}$\\
$<\xi_{2}>$ &&& $u_{2}$\\
$<\xi_{3}>$ &&& $u_{3}$\\

\hline
\hline
\end{tabular}
\caption{VEV for the scalar fields used in the model. Here,   no driving fields are introduced to the model while the minima are derived by analysing the F-terms of the flavons themselves as in Ref. \cite{Feruglio:2011qq}. With this consideration, the VEV alignment of the triplets and pentaplet use in the model can be realised from the superpotential in Eq.(\ref{w1}) and the details are shown in Appendix B}
\label{t2}
\end{table}
Since the VEVs of the scalars are realised, the mass matrix for the leptons can be calculated after spontaneous symmetry breaking. The charged lepton mass matrix obtained, has a unique structure as shown below

\begin{equation}
\label{el}
M_{l}=\begin{pmatrix}
\frac{1}{\lambda^{3}}y_{1}v_{d}v_{1}u^{2}_{1} &\frac{1}{\lambda^{2}}y_{5}v_{d} v_{1}u_{1}&0\\
\frac{1}{\lambda^{3}}y_{4}v_{d}v_{2}u_{2}u_{1}&\frac{1}{\lambda^{2}}y_{2}v_{d} v_{2}u_{2}&0\\
0&0&\frac{1}{\lambda}y_{3}v_{d}v_{3}
\end{pmatrix}=\begin{pmatrix}
m_{11}&m_{12}&0\\
m_{21}&m_{22}&0\\
0&0&m_{33}
\end{pmatrix}.
\end{equation} 

In the neutrino sector, after the electroweak symmetry breaking, the neutrino mass matrix in the basis $( \nu_{L}, N_{R}^{c}, S_{R}^{c})$ has the structure
\begin{equation}
M_{\nu}=\begin{pmatrix}
0&M_{D}&M_{L}\\
M_{D}^{T}&0&M\\
M^{T}_{L}&M^{T}&0
\end{pmatrix}.
\label{k1}
\end{equation}
After block diagonalisation, the light active neutrino mass matrix arising from the linear seesaw mechanism \cite{fo11,Hernndez2023} has the form
\begin{equation}
\label{el11}
m_{\nu}=M_{D}(M_{L}M_{R}^{-1})^{T}+M_{L}M_{R}^{-1}M_{D}^{T},
\end{equation}
where sub matrices $M_{L}$, $M_{D}$ and $M_{R}$ have the following patterns,
\begin{align}
M_{L}=m\begin{pmatrix}
1& 0& 0\\ 0&0& 1\\ 0& 1& 0
\end{pmatrix},\hspace{0.05cm} M_{D}=\left(\begin{array}{ccc}
d+\frac{2}{3}(b+c)&\frac{b}{\sqrt{2}}&\frac{b}{\sqrt{2}}\\
\frac{b}{\sqrt{2}}&c&d-\frac{1}{3}(b+c)\\
\frac{b}{\sqrt{2}}&d-\frac{1}{3}(b+c)&c
\end{array}\right)\text{and}\hspace{0.1cm} M_{R}=\begin{pmatrix}
M&0&0\\
0&0&M\\
0&M&0
\end{pmatrix}.
\label{el1}
\end{align}
where $\frac{y_{c}v_{u}u_{3}}{\lambda}=m$, $\frac{y_{a}pv_{u}}{\lambda}=b$,  $\frac{y_{a}qv_{u}}{\lambda}=c$, $\frac{y_{b}uv_{u}}{\lambda}=d$, and $M=\frac{y_{N}}{\lambda}u_{2}^{2}$, and using Eq.(\ref{el1}) in Eq.(\ref{el11}), the light active neutrino mass matrix has the following  resultant structure
\begin{equation}
m_{\nu}=\begin{pmatrix}
(4 (b + c) m)/(3 M)& (\sqrt{2} b m)/M& (\sqrt{2} b m)/M\\
 ( \sqrt{2} b m)/M& (2 c m)/M& (2 (-b - c) m)/(3 M)\\ 
 (\sqrt{2} b m)/ M& (2 (-b - c) m)/(3 M)& (2 c m)/M
\end{pmatrix}.
\label{g1}
\end{equation} 

On the other hand, the six doubly degenerate quasi-Dirac heavy neutrinos have the mass matrix structures \cite{Hernndez2023}
\begin{equation}
M_{N^{-}}=-M_{R}-M_{D}^{T}M_{D}M_{R}^{-1}+\frac{1}{2}[M_{D}M_{R}^{-1}M_{L}^{T}+M_{L}^{T}M_{R}^{-1}M_{D}^{T}].
\label{l1}
\end{equation}
\begin{equation}
M_{N^{+}}=M_{R}+M_{D}^{T}M_{D}M_{R}^{-1}+\frac{1}{2}[M_{D}M_{R}^{-1}M_{L}^{T}+M_{L}^{T}M_{R}^{-1}M_{D}^{T}].
\label{l2}
\end{equation}
So, the physical neutrino spectrum consists of three active neutrinos and six nearly degenerate quasi-Dirac heavy sterile neutrinos. Here, it can be seen that in the limit $M_{L}\rightarrow 0$, the active light neutrinos become massless. In the model, the light active neutrino masses can be explained from Eq.(\ref{g1}) by varying model parameter $M$ and  $m$. However, if we consider the cLFV processes $(l_{i}\rightarrow l_{j} \gamma)$  and $m_{\beta \beta}$ sensitivity , the value of $m$ and $M$ are found to be constrained in the specific range.

\section{ Diagonalisation} 
 The  charged leptons mass matrix obtained in Eq.(\ref{el}) can be diagonalized through the use of a rotation matrix $R_{12}$, which diagonalizes the charged lepton mass matrix as

\begin{align*}
\label{l1}
R_{12}^{\dagger}M_{L}M_{L}^{\dagger}R_{12}=R_{12}^{\dagger}hR_{12}=&\begin{pmatrix}
m_{e}^{2} &0&0\\
0&m_{\mu}^{2}&0\\
0&0&m_{\tau}^{2}
\end{pmatrix}, 
\end{align*}
\begin{align}
\text{where},~ R_{12}=\begin{pmatrix}
\cos \theta &\sin \theta e^{i \psi}&0\\
-\sin \theta e^{-i \psi}&\cos \theta&0\\
0&0&1
\end{pmatrix}
\end{align}

The neutrino mass matrix obtained in Eq.(\ref{k1}) has dimension of $9\times 9$ and it can be diagonalised by unitary matrix $\mathcal{U}$ \cite{Batra_2023}. 
On the other hand, the resultant light active neutrino mass matrix given in  Eq.(\ref{g1}) can be diagonalised by  $U_{GR1}$ as 
\begin{equation}
U_{GR1}^{T}m_{\nu}U_{GR1}=\text{diag}(m_{1},~m_{2},~m_{3}),
\end{equation} 
where, $m_{i}$, when $i$=1, 2, 3 are neutrino mass eigenvalues, and the mixing matrix $U_{GR1}$ is given by 
 \begin{equation}
 U_{GR1}=\begin{pmatrix}
 \frac{\sqrt{\phi_{g}}}{\sqrt[4]{5}}&-\frac{\sqrt{\frac{1}{\phi_{g}}}}{\sqrt[4]{5}}&0\\
 \frac{\sqrt{\frac{1}{\phi_{g}}}}{\sqrt{2}\sqrt[4]{5}}&\frac{\sqrt{\phi_{g}}}{\sqrt{2}\sqrt[4]{5}}&-\frac{1}{\sqrt{2}}\\
 \frac{\sqrt{\frac{1}{\phi_{g}}}}{\sqrt{2}\sqrt[4]{5}}&\frac{\sqrt{\phi_{g}}}{\sqrt{2}\sqrt[4]{5}}&\frac{1}{\sqrt{2}}
 \end{pmatrix},\hspace{0.5cm} \phi_{g}=\frac{1+\sqrt{5}}{2}.
 \end{equation}

 The PMNS leptonic mixing matrix $U$ for this scenario can be constructed as
 \begin{equation}
 U=R_{12}^{\dagger}U_{GR1}=\begin{pmatrix}
\frac{-\sqrt{2} e^{-i \psi} \sin \theta \sqrt{\frac{1}{\phi_{g}}} - 
 2 \cos \theta \sqrt{\phi_{g}}}{\sqrt{2}\sqrt[4]{5}}&\frac{2 \cos \theta \sqrt{\frac{1}{\phi_{g}}} - \sqrt{2} e^{-i \psi} \sin \theta \sqrt{\phi_{g}}}{\sqrt{2}\sqrt[4]{5}}&\frac{e^{i \psi} \sin \theta}{\sqrt{2}}\\
 -\frac{\sqrt{2} \cos \theta \sqrt{\frac{1}{\phi_g}} + 
 2 e^{i \psi} \sin \theta \sqrt{\phi_g}}{\sqrt{2}\sqrt[4]{5}}&\frac{2 e^{i \psi} \sin \theta \sqrt{\frac{1}{\phi_{g}}} - 
 \sqrt{2} \cos \theta \sqrt{\phi_{g}}}{\sqrt{2}\sqrt[4]{5}}&\frac{\cos \theta }{\sqrt{2}} \\
 -\frac{\sqrt{\frac{1}{\phi_{ g}}}}{\sqrt{2}\sqrt[4]{5}} &-\frac{\sqrt{\phi_{ g}}}{\sqrt{2}\sqrt[4]{5}}&-\frac{1}{\sqrt{2}}
\end{pmatrix}.
\label{eu1}
 \end{equation}
  Following the PDG convention \cite{ParticleDataGroup:2018ovx},  the standard form of the PMNS matrix U takes the form
\begin{equation}
\label{u}
 U_{\text{PMNS}} =  
	\begin{pmatrix}
	c_{12} c_{13} & s_{12} c_{13} & s_{13}e^{-i\delta} \\
	-s_{12} c_{23} - c_{12} s_{23} s_{13}e^{i\delta} & c_{12}c_{23} - s_{12} s_{23} s_{13}e^{i\delta} & s_{23} c_{13} \\
	s_{12} s_{23} - c_{12} c_{23} s_{13}e^{i\delta} & -c_{12} s_{23} - s_{12} c_{23} s_{13}e^{i\delta} & c_{23} c_{13}
	\end{pmatrix}
	\begin{matrix}
	  P
	\end{matrix},
 \end{equation}
   where $c_{ij}=\cos \theta_{ij}$ and $s_{ij}=\sin{\theta}_{ij}$ (for $\text{ij}= 12,$ 13, 23), $\delta$ is the Dirac CP violating phase, while $P=\text{diag}(e^{i\alpha/2},e^{i\beta/2},1)$ contains  two  Majorana CP  phases $\alpha$ and $\beta$. The correlation between the leptonic mixing parameters and the model parameters can be explained by comparing Eq.(\ref{eu1}) and Eq.(\ref{u}) as
\begin{equation}
\label{ep}
s^{2}_{13}=\frac{1}{2} \sin^{2} \theta,\hspace{0.4cm} s^{2}_{23}=\frac{\cos^{2} \theta}{2-\sin^{2}\theta},\hspace{0.4cm} s^{2}_{12}=\frac{\sqrt{2}\cos \psi \sin 2\theta +\phi_{g} \sin^{2} \theta + \frac{2 \cos^{2} \theta}{\phi_{g}}}{\sqrt{5}(2-\sin^{2} \theta)}.
\end{equation}
 The mixing angles $\theta_{23}$ and $\theta_{12}$ are  then related to reactor angle $\theta_{13}$ as:
\begin{equation}
\label{ea}
s^{2}_{23}=\frac{1-2 \sin^{2}\theta_{13}}{2(1-\sin^{2}\theta_{13})}\hspace{0.2cm} \text{and}~ s^{2}_{12}=\frac{2 \cos \psi \sin \theta_{13} \sqrt{1-2 \sin^{2} \theta_{13}} +\phi_{g} \sin^{2} \theta_{13} + \frac{1-2 \sin^{2} \theta_{13}}{\phi_{g}}}{\sqrt{5}(1-\sin^{2} \theta_{13})}.
\end{equation}
Furthermore, the Jarlskorg invariant $J_{CP}$, and the invariants $I_{1}$ and $I_{2}$ for this scenario can be extracted from the relations
\begin{align}
\label{ej}
J_{CP}=&\text{Im} (U_{11}^{*}U_{23}^{*}U_{13}U_{21})=\frac{-\cos \theta \sin \theta \sin \psi}{2\sqrt{10}}=\frac{-\sin \theta_{13}\sqrt{1-2\sin^{2}\theta_{13}}\sin\psi}{2\sqrt{5}}.
\end{align}
\begin{align}
\label{maj1}
I_{1}=&\text{Im}(U_{11}^{*}U_{12})=\frac{-\cos \theta \sin \theta \sin \psi}{\sqrt{2}}=-\sin \theta_{13}\sqrt{1-2\sin^{2}\theta_{13}}\sin\psi
\end{align}
\begin{align}
\label{maj2}
I_{2}=&\text{Im}(U_{11}^{*}U_{13})=\frac{\phi_{g}\cos \theta \sin \theta \sin \psi}{\sqrt{10}}=\frac{\phi_{g}\sin \theta_{13}\sqrt{1-2\sin^{2}\theta_{13}}\sin\psi}{\sqrt{5}}
\end{align}
 The Dirac CP violating phase $\delta$ can be obtained by comparing Eq.(\ref{ej}) with the definition of $J_{CP}$ in the standard parametrisation \cite{Krastev:1988yu},
\begin{equation}
J_{CP}=\frac{1}{8}\sin 2\theta_{12} \sin 2\theta_{23}\sin 2\theta_{13}\cos \theta_{12} \sin \delta.
\end{equation}
The Majorana CP phases $\alpha$ and $\beta$ can be explained by comparing Eq. (\ref{maj1}) and (\ref{maj2}) with the relations given in Ref.\cite{Jarlskog:1985ht}

\begin{equation}
I_{1}=c_{12}c_{13}^{2}s_{12}\sin{\frac{\alpha}{2}}
\end{equation}

\begin{equation}
I_{1}=c_{12}c_{13}s_{13}\sin{(\frac{\beta}{2}-\delta)}
\end{equation}
 
As shown in Eq.(\ref{eu1}), the contribution from the charged lepton sector plays a crucial role in explaining leptonic mixing parameters; in contrast, the contribution from the neutrino mass model parameters $b,~c$ and $d$ to the lepton mixing matrix is trivial. However, the information for light neutrino masses can be explained by the neutrino mass model parameters by adjusting their values to fix the mass square differences ($\Delta m_{21}^{2}$ and $\Delta m_{31}^{2}$) to their current experimental bound and the sum of the absolute neutrino masses $\sum m_{i}\leq 0.12$ eV. The sensitivity of cLFV decay processes considered in our work and the neutrinoless double beta decay parameters $m_{\beta \beta}$  give constraints on the parameter space of neutrino mass parameter $b$ and $c$. We also found that for the resultant pattern of $m_{\nu}$, no data points can simultaneously explain lepton mixing angles and masses for an inverted hierarchy case. As a result, our model only supports NH.

\section{Charged Lepton Flavor Violation}
\begin{figure}[h]
\centering
\includegraphics[scale=1]{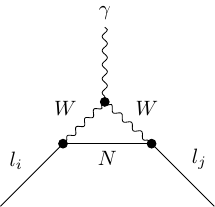}
\caption{Feynman diagram for $l_{i}\rightarrow l_{j}\gamma$ decay in seesaw mechanism}
\label{fy1}
\end{figure}

We discusses the charged lepton flavor violation (cLFV) process, especially two body  $l_{i} \rightarrow l_{j}\gamma$ decay, and its implication to the present model. In the linear seesaw scenario, the decay process is induced by exchanging the nine neutrinos (consisting of three active light neutrinos and six heavy neutrinos) coupled to the charged leptons in the charged current (CC) \cite{fo11}. The Feynman diagram for this contribution is given in Fig \ref{fy1}. The resultant decay ratio for the $l_{i} \rightarrow \l_{j}\gamma$ is given by \cite{fo11}

\begin{align}
Br~(l_{i} \rightarrow l_{j}\gamma)&=\frac{\alpha^{3}_{W}s^{2}_{W}}{256\pi^{2}}\frac{m_{i}^{5}}{M_{W}^{4}}\frac{1}{\Gamma_{i}}|G_{ij}^{W}|^{2},
\end{align}
\begin{align*}
\text{where}, \hspace{0.2cm} G_{ij}^{W}&=\sum^{9}_{k=1} K^{*}_{ik}K_{jk}G_{\gamma}^{W}(\frac{m_{N_{K}}}{M_{W}^{2}}), \hspace{0.2cm} \text{and}
\end{align*}
\begin{equation}
G_{\gamma}^{W}(x)=\frac{1}{12(1-x)^{4}}(10-43x+78x^{2}-49x^{3}+18x^{3}\ln{x} +4x^{4}).
\end{equation}
The Lagrangian  that describes the CC weak interaction of mass-eigenstate neutrinos in any seesaw model is given below
\begin{equation}
-L_{CC}=\frac{g}{2\sqrt{2}}\sum_{\alpha =1}^3 \sum_{k =1}^9 K_{\alpha k}\overline{l}_{\alpha}\gamma_{\mu}(1-\gamma_{5})N_{k}W^{\mu}+\text{h.c.,} 
\end{equation}
where, $\alpha=$1, 2, 3 denote the charged leptons and $k$ represent the three light active neutrinos with $k$ = 1, 2, 3 as well as the six mediators, with k = 4, 5, . . . 9.
The matrix $K$ is the effective lepton mixing matrix that describes the CC weak interaction. It can be expressed in rectangular form and has dimensions $3\times 9$.
 Here, K can be written as \cite{fo11}
\begin{equation}
K_{\alpha k}=\sum^{n}_{c=1}\Omega^{\dagger}_{c \alpha}\mathcal{U}_{c k}
\label{p1}
\end{equation}

where $\Omega$ is $3\times3$ unitary matrix that diagonalises the charged lepton mass matrix  while $\mathcal{U}$ diagonalises higher dimensional neutrino mass matrix. The matrix $K$ can be expressed as     
\begin{equation}
K=(K_{L},K_{H})
\end{equation} 
where $K_L$ and $K_H$ are $3 \times 3$ and $3 \times 6$ matrices, respectively. Since, the charged lepton mass matrix is non-diagonal and it is diagonalised by $R_{12}$ in our model, the contribution from this sector is prominent in $K$.  So, using $\mathcal{U}$ given in Ref.\cite{Batra_2023} \footnote{$\mathcal{U}$ is the unitary matrix that diagonalises the neutrino mass matrix given in Eq. (\ref{k1}) and its details are given in Ref. \cite{Batra_2023,fo11}} and $\Omega\simeq R_{12}$ in Eq.(\ref{p1}), we have 
\begin{equation}
K=\begin{pmatrix}
R_{12}^{\dagger}&0&0\\
0&0&0\\
0&0&0
\end{pmatrix} \begin{pmatrix}
(1-\frac{1}{2} \epsilon)U_{GR1}&-\frac{i}{\sqrt{2}}V&\frac{1}{\sqrt{2}}V\\
0&\frac{i}{\sqrt{2}}(1-\frac{1}{2}\epsilon^{'})&\frac{1}{\sqrt{2}}(1-\frac{1}{2}\epsilon^{'})\\-V^{\dagger}&-\frac{i}{\sqrt{2}}(1-\frac{1}{2}\epsilon^{'})&\frac{1}{\sqrt{2}}(1-\frac{1}{2}\epsilon^{'})
\end{pmatrix}
\end{equation} 

Here, $V$ is $3\times 3$ matrix and $\epsilon$ and $\epsilon^{'}$ denote non-unitary corrections. Their explicit form is denoted by\\
\begin{equation}
V=M_{D}(M_{R}^{-1})^{T}
\end{equation}
\begin{equation}
\epsilon \approx (M_{D}M_{R}^{-1 T})(M_{R}^{-1}M_{D}^{\dagger})+\mathcal{O}(M_{L}^{2}/M_{R}^{2})
\end{equation}
\begin{equation}
\epsilon^{'} \approx (M_{R}^{-1 }M_{D}^{\dagger})(M_{D}M_{R}^{-1T})+\mathcal{O}(M_{L}^{2}/M_{R}^{2})
\end{equation}

Then, the sub-matrices $K_{L}$ and $K_{H}$ have the forms:
\begin{align}
K_L&=R^{\dagger}_{12}(I_{3\times 3}-\frac{1}{2}M_{D}(M_{R}^{-1})^{T}M_{R}^{-1}M_{D}^{\dagger})
U_{GR1}=R^{\dagger}_{12}(I_{3\times 3}-\frac{1}{2}VV^{\dagger})U_{GR1}\nonumber\\
&=R^{\dagger}_{12}(I_{3\times 3}-\frac{1}{2}\eta)U_{GR1}, ~\text{where}~ \eta=VV^{\dagger},
\end{align}
\begin{equation}
K_{L}\simeq U
\end{equation}
\begin{align}
K_H=(-\frac{i}{\sqrt{2}}R^{\dagger}_{12}V,\frac{i}{\sqrt{2}}R^{\dagger}_{12}V).
\end{align}

The term $\eta$ characterises the deviation from the  non-unitarity of the light neutrino mixing matrix. The upper limit on the non-unitarity of the leptonic mixing matrix is given in Ref \cite{fernandez2016global}. In the low scale seesaw mechanism such as linear seesaw mechanism the value of $\eta$ can be relatively large \cite{Batra_2023}. Here,  the   $l_{i}\rightarrow l_{j}\gamma$ decay process is due to the exchange of the three active neutrinos and six sub-dominantly coupled heavy quasi-Dirac neutrinos.  In this case, the CC contribution is sizeable even in the absence of supersymmetry (SUSY).  It is to be noted that in supersymmetry, cLFV processes are induced by the misalignments of sleptons and charged lepton mass matrices which is often introduced to the theory radiatively \cite{Masina:2005am, Hisano:2001qz}. In this work, we discuss two body cLFV processes, $\mu \rightarrow e+\gamma$,  $\tau \rightarrow e+\gamma$ and $\tau \rightarrow \mu +\gamma$ and study their implication on our model without considering SUSY contribution and radiative correction. 

The other cLFV violating processes such as three body decay ($l_{i}\rightarrow l_{j}l_{k}l_{k}$) and the processes involving muonic atom($\mu \rightarrow e$ conversion in nuclei \cite{Gautam:2020bnx, Abada:2015oba}, $\mu^{-} e^{-}\rightarrow e^{-} e^{-} $\cite{Koike:2010xr} ) are discussed with the additional Feynman diagrams and their loop functions in Ref.[\cite{Abada:2015oba, Ilakovac:1994kj}]. Following the detailed calculations of the branching ratio and conversion ratio using Ref.\cite{Gautam:2020bnx,Abada:2015oba}, it is difficult to find the parameter space that can explain the current sensitivity of $m_{\beta \beta}$ and the future and current sensitivity of the the respective cLFV decays simultaneously in our numerical analysis. These less predictive results may be attributed from the lesser restriction on model parameter $m$ from the current and future sensitivity of branching ratio and conversion ratio in contrast to the two body decay discussed in Section \ref{Re}.  Hence, the results of these cLFV  processes are not presented  in this work.
\subsection{$\mu \rightarrow e+\gamma$ decay}
The $\mu \rightarrow e+\gamma$ decay is the most sensitive of cLFV processes. This decay process is aimed to be investigated by MEG experiment\cite{PhysRevLett.110.201801}, and the future planned more sensitive MEG II \cite{meucci2022meg}. In the linear seesaw framework, the decay ratio can be written as
\begin{align}
Br~(\mu \rightarrow e\gamma)&=\frac{\alpha^{3}_{W}s^{2}_{W}}{256\pi^{2}}\frac{m_{\mu}^{5}}{M_{W}^{4}}\frac{1}{\Gamma_{\mu}}|G_{\mu e}^{W}|^{2},
\end{align} 
where, $\Gamma_{\mu}$ is the total decay width of muon. The current sensitivity and the future sensitivity of this decay is shown in Table. \ref{p4}.

\subsection{$\tau \rightarrow e+\gamma$ and $\tau \rightarrow \mu+\gamma$ decays}
 The tau lepton decays have many flavor-violating channels, but their amplitude is enhanced by many orders of magnitude in comparison with muon decays. Here, we investigate both the two body $\tau$ decay namely  $\tau \rightarrow e+\gamma$ and $\tau \rightarrow \mu+\gamma$ decay processes. The branching ratio of these processes are given by 
\begin{align}
Br~(\tau \rightarrow e\gamma)&=\frac{\alpha^{3}_{W}s^{2}_{W}}{256\pi^{2}}\frac{m_{\tau}^{5}}{M_{W}^{4}}\frac{1}{\Gamma_{\tau}}|G_{\tau e}^{W}|^{2},
\end{align}
\begin{align}
Br~(\tau \rightarrow e\gamma)&=\frac{\alpha^{3}_{W}s^{2}_{W}}{256\pi^{2}}\frac{m_{\tau}^{5}}{M_{W}^{4}}\frac{1}{\Gamma_{\tau}}|G_{\tau \mu}^{W}|^{2},
\end{align}  
     where, $\Gamma_{\tau}=2.1581 \times 10^{-12}$ GeV \cite{Awasthi:2013we} is the total decay width of tau lepton. Experiments like BaBar \citep{MIYAZAKI2011251} and Belle \cite{Benes:2005hn} provide limits to tau lepton decay processes. 
     
\begin{table}[h]
\begin{center}    
\begin{tabular}{ccc}
\hline 
\hline
cLFV process & ~~~~Present bound &~~~~ Future Sensitivity \\ 
\hline 
\hline
\vspace{0.3cm}
$\mu \rightarrow e \gamma $ & 5.7$\times 10^{-13}$ & 6.0$\times 10^{-14}$ \\  
$\tau \rightarrow e \gamma $ & 3.3$\times 10^{-8}$ & $\sim 3 \times 10^{-9}$ \\
$\tau \rightarrow \mu \gamma $ & 4.4$\times 10^{-8}$ & $\sim  10^{-9}$ \\
\hline 
\hline
\end{tabular}
\caption{\footnotesize{Present and future sensitivity for the cLFV processes \cite{Gautam:2020bnx} consider in our work.}}
\label{p4}
\end{center}
\end{table}
 
\section{Neutrinoless Double Beta Decay}
 
In the linear seesaw framework, there are two separate contributions to $0\nu \beta \beta$ transition due to light and heavy neutrino exchanges, just as in the inverse seesaw framework mentioned in \cite{Awasthi:2013we}. Here, we will investigate the effect on effective neutrino mass $m_{\beta \beta}$ due to these heavy neutrinos. The effective Majorana mass $m_{\beta \beta}$ in standard case considering three active light neutrinos is given by
\begin{equation}
m_{\beta \beta}=|\sum_{i}^{3}U_{ei}^{2}m_{i}|
\label{m1}
\end{equation}
 
 The upper limits of $m_{\beta \beta}$ are provided  by several experiments such as GERDA \cite{GERDA:2019ivs} with $m_{\beta \beta}$ $<(104 -228)$ meV corresponds to $^{76}Ge~(T^{0\beta \beta}_{1/2}>9 \times 10^{25}$yr), CUORE \cite{CUORE:2019yfd} with  $m_{\beta \beta} <(75 -350)$ meV corresponds to $^{130}Te~(T^{0\beta \beta}_{1/2}>3.2 \times 10^{25}$yr) with KamLAND-Zen \cite{KamLAND-Zen:2016pfg} as $m_{\beta \beta} <(61 -165)$ meV corresponds to $^{136}Xe~(T^{0\beta \beta}_{1/2}>1.07\times 10^{27}$ yr).
 
  The presence of six additional heavy neutrinos modify Eq.(\ref{m1}) into the following \cite{Awasthi:2013we}
\begin{equation}
m_{\beta \beta}=U_{ei}^2 m_{i}+\mathcal{N}_{ej}\frac{\mathcal{M}_{j}}{p^{2}-\mathcal{M}_{j}^2}|<p>|^2 ,
\end{equation}
  where $i$=1,2,3 and $j$= 4,5,...,9. Here, $\mathcal{N}$ is the active sterile mixing matrix $K_{H}$, $\mathcal{M}_{j}$ are the mass of the six quasi-Dirac heavy neutrinos, and $|<p>| \simeq$ 190 MeV represents the neutrino virtuality momentum. 
\section{Results and Discussion \label{Re}} 
\begin{figure}[h]
\includegraphics[width=.5\textwidth]{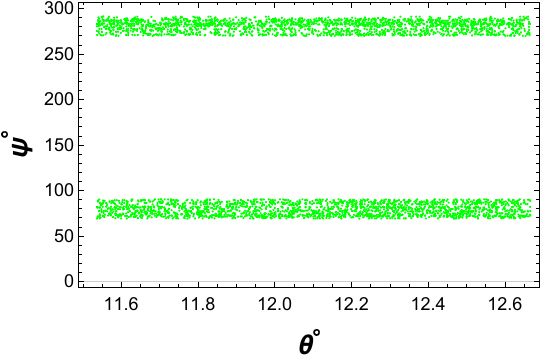}
\caption{Allowed parameter space of model parameters $\theta$ and $\psi$ for NH.}
\label{r1}
\end{figure}

\begin{figure}[h]
\subfigure[]{
\includegraphics[width=.45\textwidth]{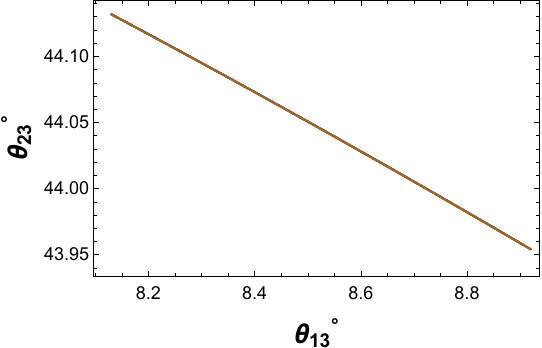}}
\quad
\subfigure[]{
\includegraphics[width=.45\textwidth]{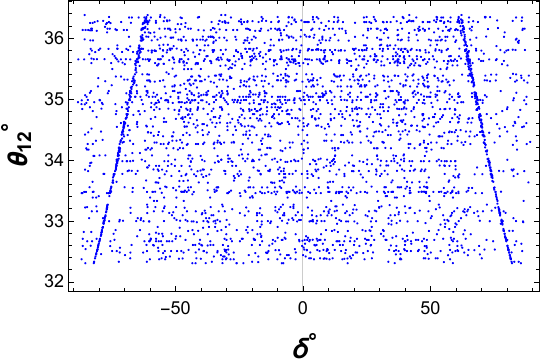}}
\subfigure[]{
\includegraphics[width=.45\textwidth]{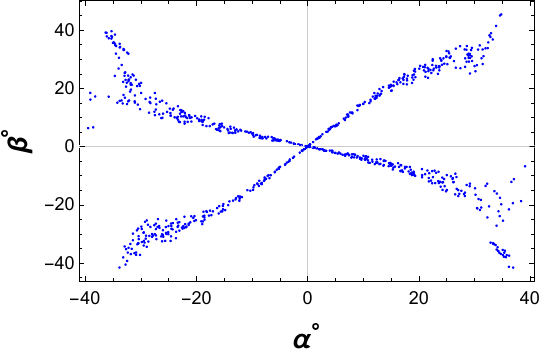}}
\caption{Correlation plots between different neutrino  parameters for NH.}
\label{r2}
\end{figure}

In section III, it is shown in Eq.(\ref{ea}) that  there is correlation between the reactor angle $\theta_{13}$ and other two mixing angles $(\theta_{23}~ \text{and}~\theta_{12})$.   So, the values of the $\theta_{23}~ \text{and}~\theta_{12}$ can be predicted by using the values of $\theta_{13}$ in $3\sigma$  range of the current global fit data as input. Additionally, the  rotational matrix  parameter $\psi$ which is necessary for explaining the mixing angle $\theta_{12}$ , is taken in the range $(0,~ 360^{\circ})$. After evaluating and selecting the values of $\theta_{12}$ that lie in the current experimental bound, the values of $\psi$  are found to be in a definitive range, whereas the range for the rotational matrix parameter $\theta$ can be extracted from Eq.(\ref{ep}). The allowed region of $\psi$ that can satisfy current data is shown in Fig.\ref{r1}.  The model predictions for $\theta_{23}$, $\theta_{12}$,  $\delta$, $\alpha$ and $\beta$ in the normal hierarchy (NH) case are shown in the correlation plots Fig.\ref{r2}. From Fig.\ref{r2}, it is evident that the present model successfully predicts $\theta_{23}$ values in the lower octant range ($<45^{\circ}$), which lie well within the 3$\sigma$ range of the current oscillation data. This outcome is a significant characteristic of our lepton mass model. However, the values of $\theta_{12}$ are dispersed throughout the $3{\sigma}$ range. This is due to the relaxation of the constraint imposed on the parameter $\psi$. The Dirac CP-violating phase $\delta$ is also obtained in the 3$\sigma$ range, as shown in Fig.\ref{r2}. Furthermore, the two Majorana CP phases are found to be constrained in the specific range as shown in Fig.\ref{r2}(c).  Thus, the contribution from the specific charged lepton mass matrix predicted by our model, can give necessary deviations to exact GR1 mixing, and these deviations are required to explain current lepton mixing data.

 \begin{figure}[h]
\subfigure[]{
\includegraphics[width=.45\textwidth]{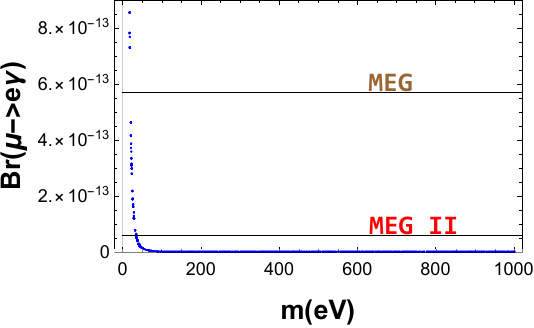}}
\quad
\subfigure[]{
\includegraphics[width=.45\textwidth]{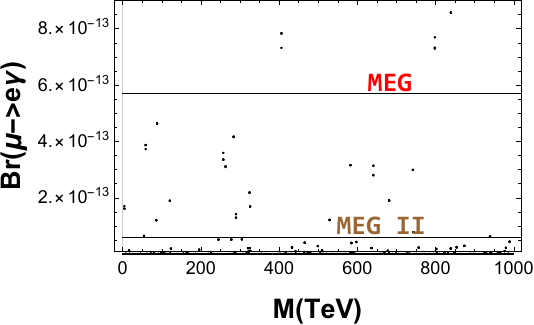}}
\caption{Correlation plots between (a) $m$ and $Br(\mu \rightarrow e \gamma)$ and (b) $M$ and $Br(\mu \rightarrow e \gamma)$.}
\label{mbr}
\end{figure}

\begin{figure}[h]
\subfigure[]{
\includegraphics[width=.45\textwidth]{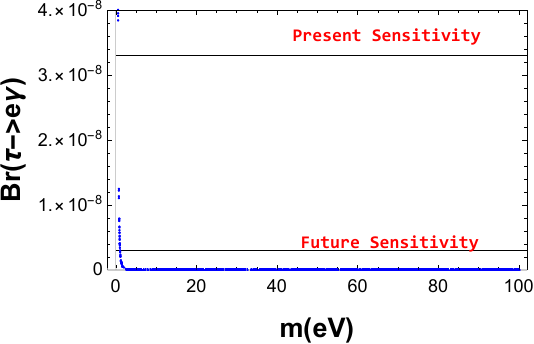}}
\quad
\subfigure[]{
\includegraphics[width=.45\textwidth]{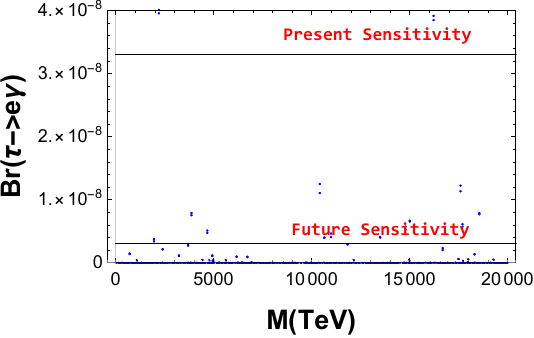}}
\caption{Correlation plots between (a) $m$ and $Br(\tau \rightarrow e \gamma)$ and (b) $M$ and $Br(\tau \rightarrow e \gamma)$.}
\label{mt}
\end{figure}

\begin{figure}[h]
\subfigure[]{
\includegraphics[width=.45\textwidth]{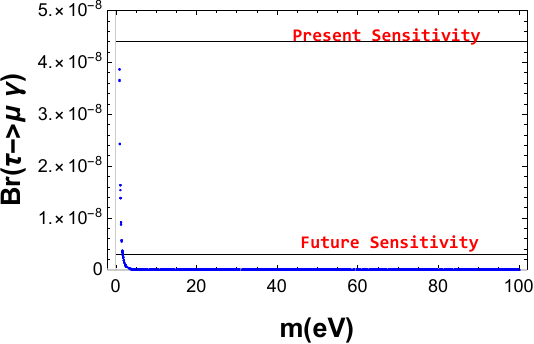}}
\quad
\subfigure[]{
\includegraphics[width=.45\textwidth]{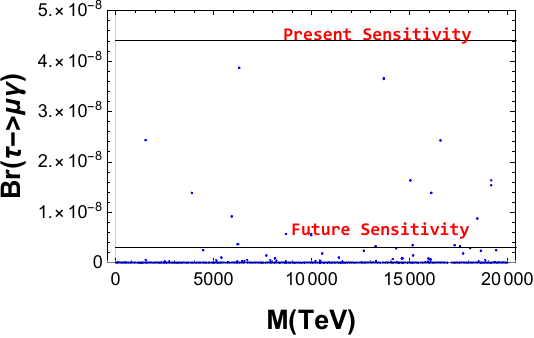}}
\caption{Correlation plots between (a) $m$ and $Br(\tau \rightarrow \mu \gamma)$ and (b) $M$ and $Br(\tau \rightarrow \mu \gamma)$.}
\label{mta}
\end{figure}

\begin{figure}[h]
\subfigure[]{
\includegraphics[width=.45\textwidth]{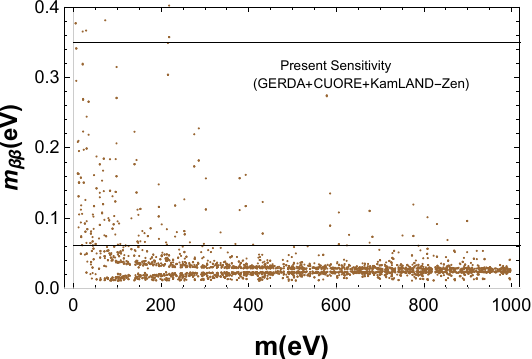}}
\quad
\subfigure[]{
\includegraphics[width=.45\textwidth]{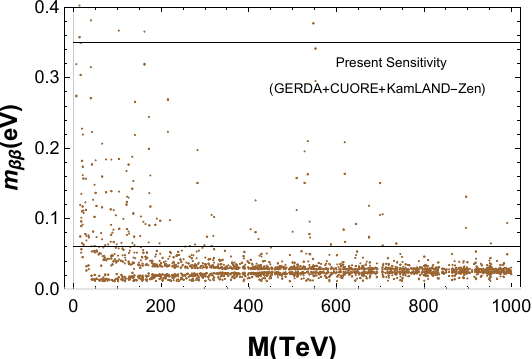}}
\caption{Variations of model parameters (a) $m$ with $m_{\beta \beta}$ and (b) $M$ with $m_{\beta \beta}$ for cases where current and future sensitivity for $Br(\mu \rightarrow e \gamma)$ is considered.}
\label{mbbu}
\end{figure}

\begin{figure}[h]
\subfigure[]{
\includegraphics[width=.45\textwidth]{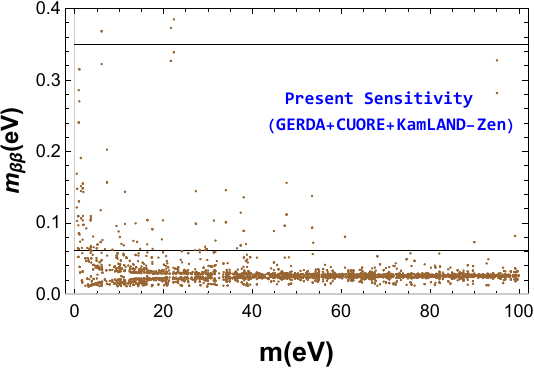}}
\quad
\subfigure[]{
\includegraphics[width=.45\textwidth]{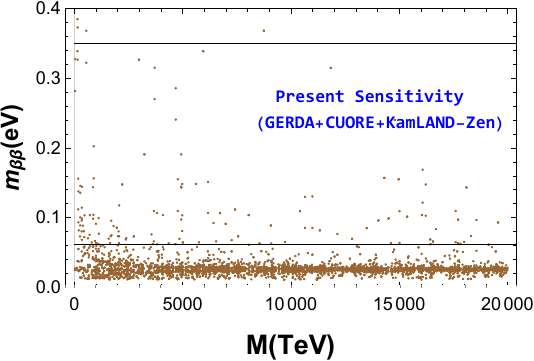}}
\caption{Variations of model parameters (a) $m$ with $m_{\beta \beta}$ and (b) $M$ with $m_{\beta \beta}$ for cases where current and future sensitivity for $Br(\tau \rightarrow e \gamma)$ is considered.}
\label{mbbt}
\end{figure}

\begin{figure}[h]
\subfigure[]{
\includegraphics[width=.45\textwidth]{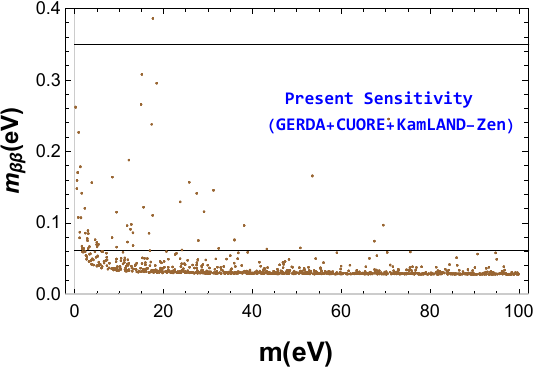}}
\quad
\subfigure[]{
\includegraphics[width=.45\textwidth]{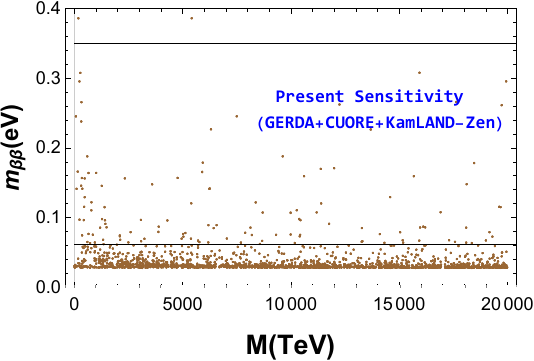}}
\caption{Variations of model parameters (a) $m$ with $m_{\beta \beta}$ and (b) $M$ with $m_{\beta \beta}$ for cases where current and future sensitivity for $Br(\tau \rightarrow \mu \gamma)$ is considered.}
\label{mbbta}
\end{figure}

\begin{figure}[h]
\subfigure[]{
\includegraphics[width=.45\textwidth]{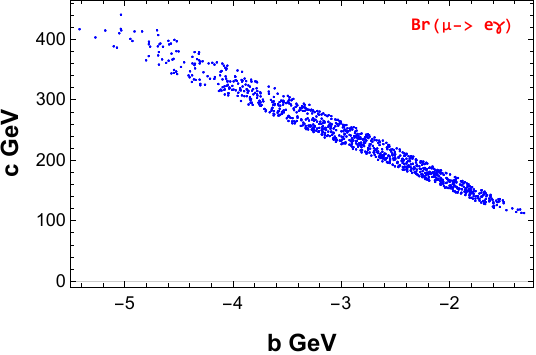}}
\quad
\subfigure[]{
\includegraphics[width=.45\textwidth]{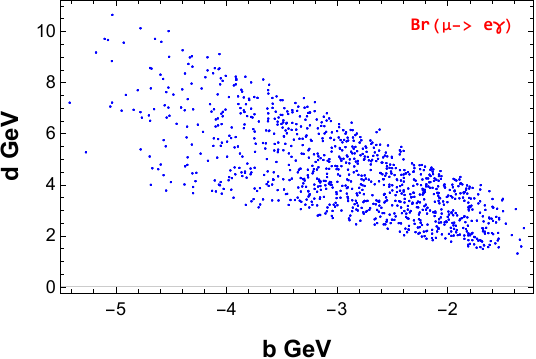}}
\subfigure[]{
\includegraphics[width=.45\textwidth]{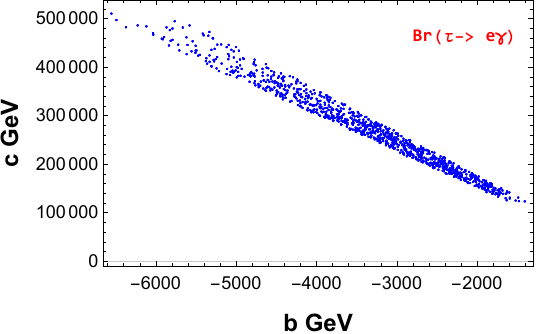}}
\quad
\subfigure[]{
\includegraphics[width=.45\textwidth]{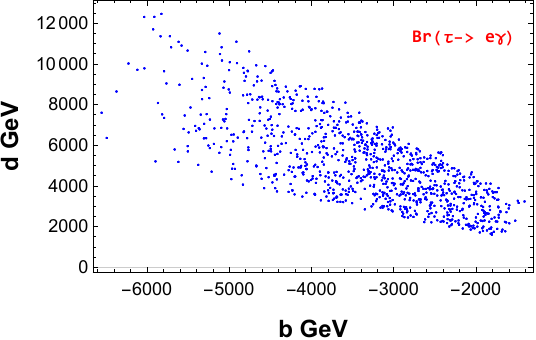}}
\caption{Allowed regions for model parameters $b$, $c$ and $d$ which can  satisfy the current neutrino masses data and present sensitivity of $m_{\beta \beta}$  along with either  $Br(\mu \rightarrow e \gamma)$ (both present and projected future sensitivity ) or $Br(\tau \rightarrow e \gamma)$ ( both present and future sensitivity).}
\label{b1}
\end{figure}

\begin{figure}[h]
\subfigure[]{
\includegraphics[width=.45\textwidth]{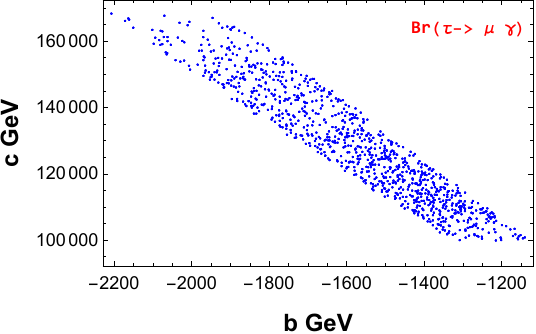}}
\quad
\subfigure[]{
\includegraphics[width=.45\textwidth]{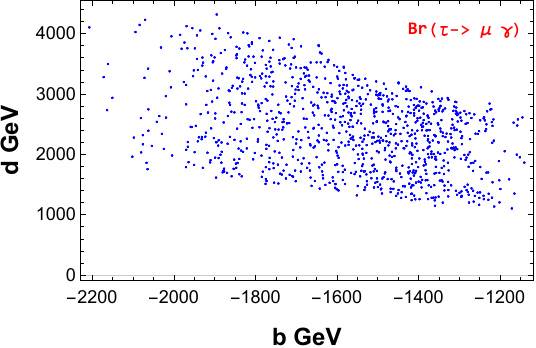}}
\caption{Allowed regions for model parameters $b$, $c$ and $d$ which can  satisfy the current neutrino masses data and present sensitivity of $m_{\beta \beta}$  along with the future sensitivity for $Br(\tau \rightarrow \mu \gamma)$.}
\label{b1a}
\end{figure}

\begin{figure}[h]
\subfigure[]{
\includegraphics[width=.45\textwidth]{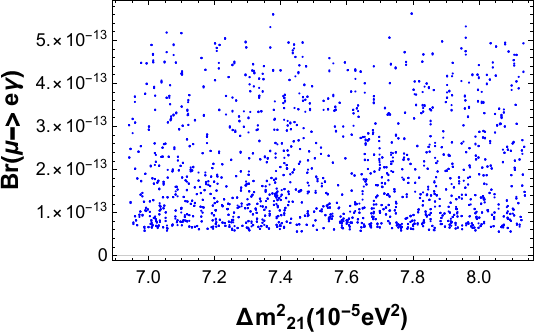}}
\quad
\subfigure[]{
\includegraphics[width=.45\textwidth]{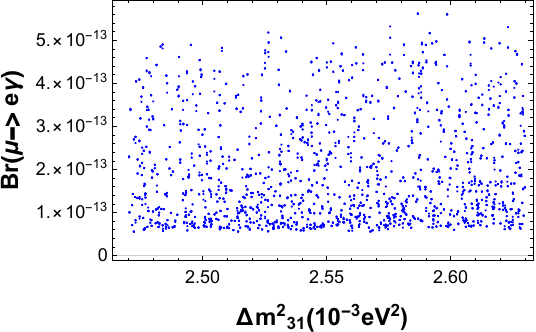}}
\caption{Correlation plots between between  (a). $\Delta m_{21}^{2}$ and  the present and future sensitivity of $Br(\mu \rightarrow e \gamma)$, and (b).$\Delta m_{31}^{2}$ and  the present and future sensitivity of $Br(\mu \rightarrow e \gamma)$.}
\label{bu1a}
\end{figure}

\begin{figure}[h]
\subfigure[]{
\includegraphics[width=.45\textwidth]{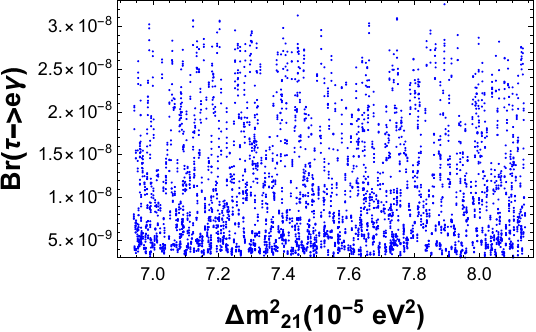}}
\quad
\subfigure[]{
\includegraphics[width=.45\textwidth]{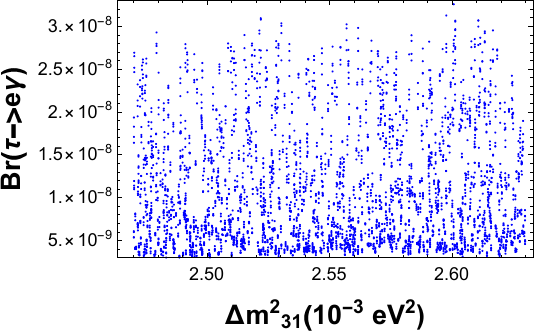}}
\caption{Correlation plots between between (a). $\Delta m_{21}^{2}$ and  the present and future sensitivity of $Br(\tau \rightarrow e \gamma)$, and (b).$\Delta m_{31}^{2}$ and  the present and future sensitivity of $Br(\tau \rightarrow e \gamma)$.}
\label{bt1a}
\end{figure}

\begin{figure}[h]
\subfigure[]{
\includegraphics[width=.45\textwidth]{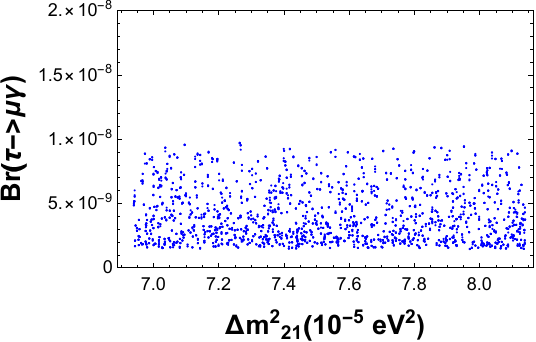}}
\quad
\subfigure[]{
\includegraphics[width=.45\textwidth]{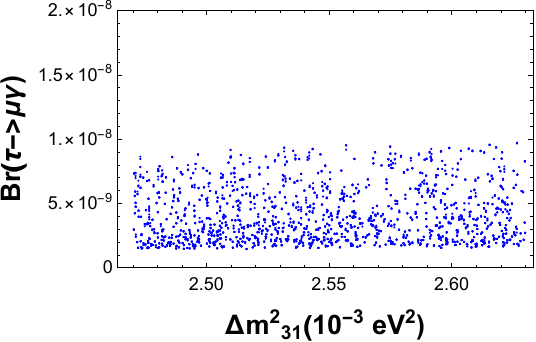}}
\caption{Correlation plots between between (a). $\Delta m_{21}^{2}$ and  the future sensitivity of $Br(\tau \rightarrow \mu \gamma)$, and (b).$\Delta m_{31}^{2}$ and  the future sensitivity of $Br(\tau \rightarrow \mu \gamma)$.}
\label{bt1b}
\end{figure}

\begin{figure}[h]
\subfigure[]{
\includegraphics[width=.45\textwidth]{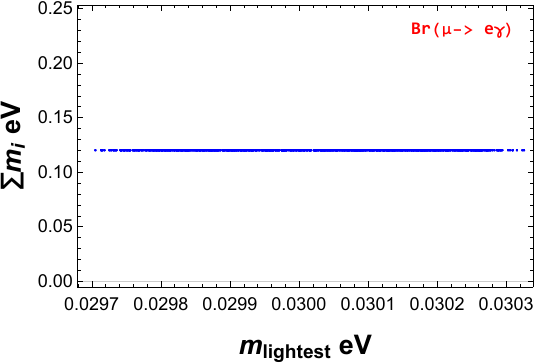}}
\quad
\subfigure[]{
\includegraphics[width=.45\textwidth]{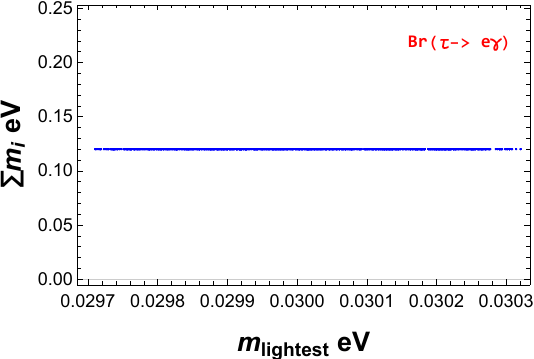}}
\subfigure[]{
\includegraphics[width=.45\textwidth]{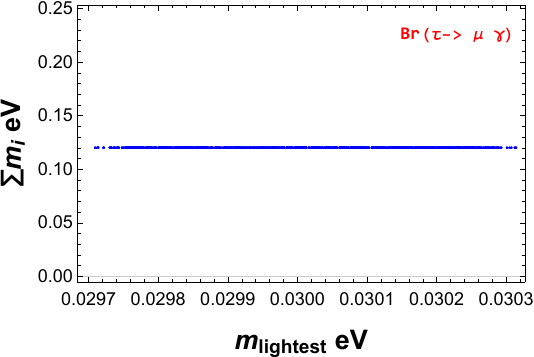}}
\caption{Correlation plots between between $m_{lightest}$ and $\sum m_{i}$ for the case where (a). the present and future sensitivity of $Br(\mu \rightarrow e \gamma)$ are considered,  (b).  the present and future sensitivity of $Br(\tau \rightarrow e \gamma)$ are considered, and (c).  the future sensitivity of $Br(\tau \rightarrow \mu \gamma)$ are considered. }
\label{sm}
\end{figure}

\begin{figure}[h]
\subfigure[]{
\includegraphics[width=.45\textwidth]{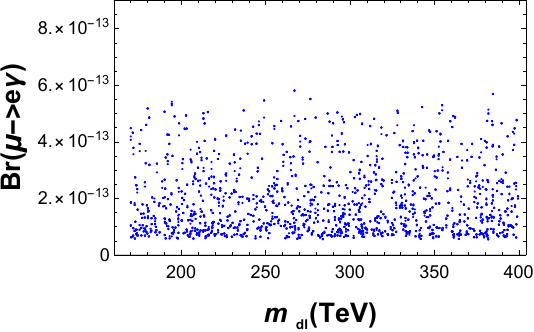}}
\quad
\subfigure[]{
\includegraphics[width=.45\textwidth]{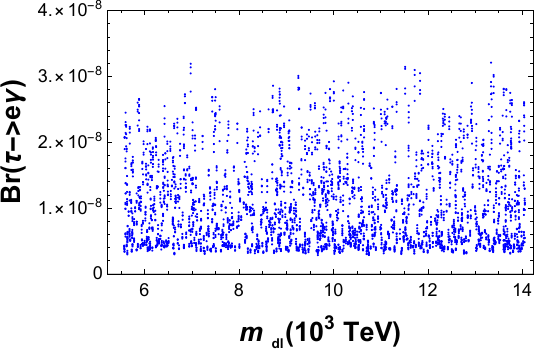}}
\subfigure[]{
\includegraphics[width=.45\textwidth]{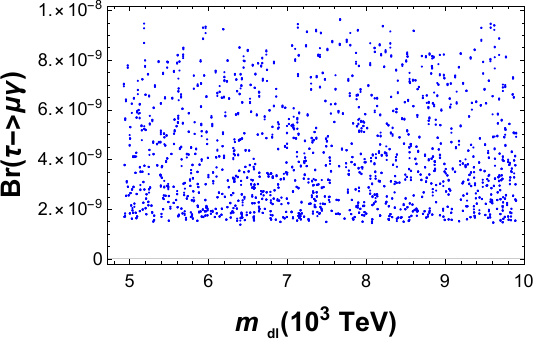}}
\caption{Correlation plots between between (a). lightest quasi-Dirac neutrino mass ($m_{dl}$) and  the present and future sensitivity of $Br(\mu \rightarrow e \gamma)$, (b).lightest quasi-Dirac neutrino mass ($m_{dl}$) and  the present and future sensitivity of $Br(\tau \rightarrow e \gamma)$, (c).lightest quasi-Dirac neutrino mass ($m_{dl}$) and  the future sensitivity of $Br(\tau \rightarrow \mu \gamma)$.}
\label{dd}
\end{figure}

 \begin{figure}[h]
\subfigure[]{
\includegraphics[width=.45\textwidth]{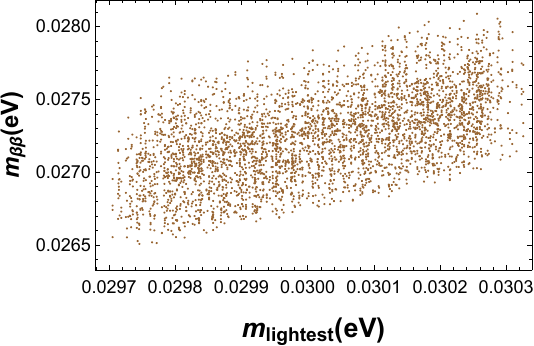}}
\quad
\subfigure[]{
\includegraphics[width=.45\textwidth]{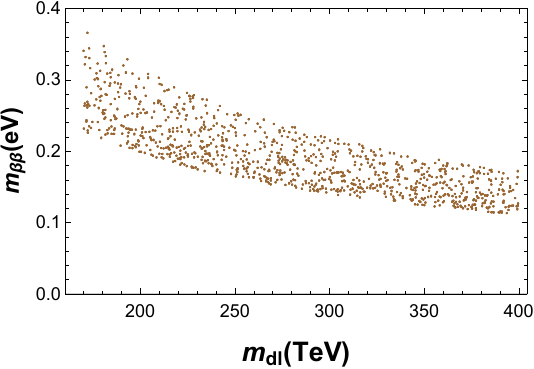}}
\caption{Correlation plots between (a) $m_{lightest}$ and $m_{\beta \beta}$ without contribution from heavy neutrinos and (b) lightest quasi-Dirac neutrinos ($m_{dl})$ and $m_{\beta \beta}$ with contribution from heavy neutrinos, for the case where the sensitivity of $\mu \rightarrow e \gamma$ is considered.}
\label{mbu1}
\end{figure}

 \begin{figure}[h]
\subfigure[]{
\includegraphics[width=.45\textwidth]{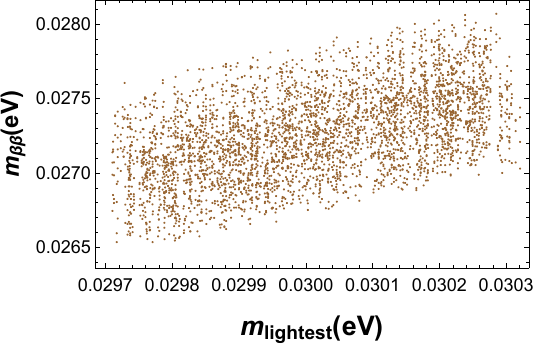}}
\quad
\subfigure[]{
\includegraphics[width=.45\textwidth]{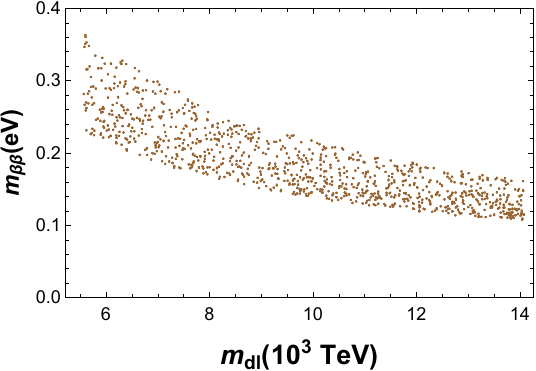}}
\caption{Correlation plot between (a) $m_{lightest}$ and $m_{\beta \beta}$ without contribution from heavy neutrinos and (b) lightest quasi-Dirac neutrinos ($m_{dl})$ and $m_{\beta \beta}$, for the case where the sensitivity of $\tau \rightarrow e \gamma$ is considered.}
\label{mbt1}
\end{figure}

\begin{figure}[h]
\subfigure[]{
\includegraphics[width=.45\textwidth]{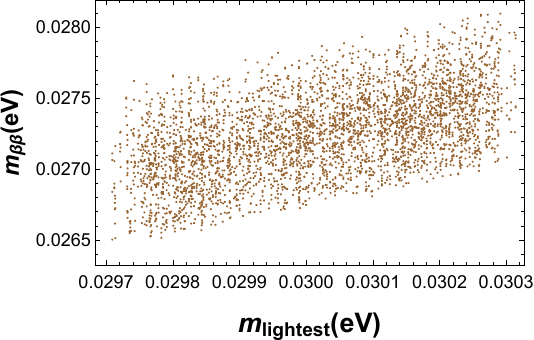}}
\quad
\subfigure[]{
\includegraphics[width=.45\textwidth]{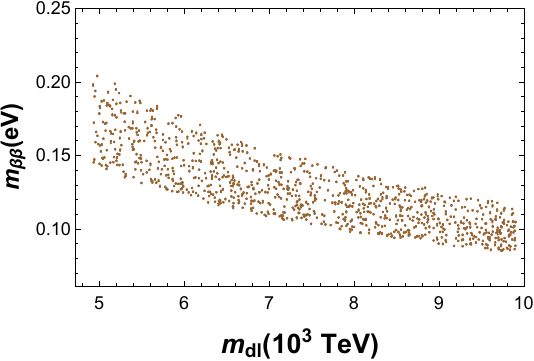}}
\caption{Correlation plot between (a) $m_{lightest}$ and $m_{\beta \beta}$ without contribution from heavy neutrinos and (b) lightest quasi-Dirac neutrinos ($m_{dl})$ and $m_{\beta \beta}$, for the case where the sensitivity of $\tau \rightarrow \mu \gamma$ is considered.}
\label{mbt2}
\end{figure}

  We also analyse the implication of  $l_{i} \rightarrow l_{j}\gamma$ decay processes on our model. In our work, the neutrino mass model parameters depend on the neutrino oscillation parameters, the cLFV decay rates, and $m_{\beta \beta}$ sensitivity. Here, to explain the the cLFV decays in their current and projected sensitivity, we vary the mass parameter $M$ in TeV ranges and $m$ in eV ranges for the all the three cases as shown in Fig.\ref{mbr}, Fig.\ref{mt} and Fig.\ref{mta}. Here, the mass parameter $m$ values are found to be constrained in a particular range for the $\mu \rightarrow e\gamma$,   $\tau \rightarrow e\gamma$ and $\tau \rightarrow \mu\gamma$ decay as shown in Fig.\ref{mbr}(a), Fig.\ref{mt}(a) and  Fig.\ref{mt}(b), respectively. However, the contribution from parameter $M$ is less restricted in all the decay processes, as shown in Fig.\ref{mbr}(b),  Fig.\ref{mt}(b) and Fig.\ref{mta}(b). 
 
  Furthermore, we analyze the implication of effective Majorana mass  $m_{\beta \beta}$ sensitivity on our model. In Fig.\ref{mbbu}, we show the variation of mass parameters $M$ and $m$ with $m_{\beta \beta}$ in the premise where muon decay rate is considered, whereas Fig.\ref{mbbt} and Fig.\ref{mbbta} show the variation of mass parameters $M$ and $m$ with $m_{\beta \beta}$ in the case where tau decay rates are considered. After considering all the current sensitivity of the processes studied in our work, Fig.\ref{b1} shows the allowed parameter space for $b$, $c$ and $d$ that can satisfy all the current neutrino oscillation parameters, $m_{\beta \beta}$ sensitivity or either $Br(\mu \rightarrow e+\gamma)$ sensitivity (both current and future) or  $Br(\tau \rightarrow e+\gamma)$ sensitivity. The model cannot predicts the allowed parameter space that can satisfy both the current sensitivity of $m_{\beta \beta}$ and $\tau \rightarrow \mu \gamma$ simultaneously. Fig.\ref{b1a} shows the allowed parameter space  that can explained the future sensitivity of $\tau \rightarrow \mu \gamma$ along with the current data for neutrino masses and current sensitivity of $m_{\beta \beta}$. As a result,   the mass square differences $\Delta m_{21}^{2}$, $\Delta m_{21}^{2}$ are also sensitive to cLFV processes, as shown in Fig.\ref{bu1a}, Fig.\ref{bt1a} and  Fig.\ref{bt1b}.
  
    Fig.\ref{sm} shows the correlation plots between the lightest active neutrino mass $m_{lightest}$ and  $\sum m_{i}$ in the premise where either one of the cLFV decay processes are taken into account. The predicted value for $\Delta m_{21}^{2}$, $\Delta m_{21}^{2}$ and  $\sum m_{i}$ are found in their current experimental bounds. Interestingly, for our model, it is found that the values of the quasi-Dirac neutrinos need to be in the TeV range to explain all the two body cLFV decay processes and current sensitivity of $m_{\beta \beta}$ as shown in Fig.\ref{dd}, Fig\ref{mbu1}(b), Fig\ref{mbt1}(b) and Fig\ref{mbt2}(b) . More specifically, the lightest quasi-Dirac neutrino ($m_{dl}$) needs to be in $\sim(1.7-4)10^{2}$ TeV range to explain both the current and future sensitivity of $Br(\mu \rightarrow e \gamma)$ and the present sensitivity of $m_{\beta \beta}$ whereas $m_{dl}$ restricts in the range $\sim (5.56-14.08 )10^{3}$ TeV, in the case where both the current and future sensitivity of $Br(\tau \rightarrow e \gamma)$ and the present sensitivity of $m_{\beta \beta}$ are considered. On the other hand, the future sensitivity of $\tau \rightarrow \mu \gamma$ and the current sensitivity of $m_{\beta \beta}$ can be accommodated when the $m_{dl}$ is in $\sim (4.92-9.90 )10^{3}$ TeV range.  
 
   Here, we find that in the absence of contribution from pseudo-Dirac sterile neutrinos, the value of $m_{\beta \beta}$ remains the same for both the allowed parameter space of $b$, $c$, and  $d$ as shown in Fig.\ref{mbu1}(a),  Fig.\ref{mbt1}(a) and Fig.\ref{mbt2}(a) . However, $m_{\beta \beta}$ varies significantly for all the cases when pseudo-Dirac sterile neutrinos contribution is considered, as shown in Fig.\ref{mbu1}(b),  Fig.\ref{mbt1}(b) and Fig.\ref{mbt2}(b) . The range of $m_{\beta \beta}$ for the $\tau \rightarrow \mu \gamma$ case is relatively lower than the other two decay cases but it is well within the current sensitivity range as shown in Fig.\ref{mbt2}(b). Fig.\ref{mbu1}(b), Fig.\ref{mbt1}(b) and Fig.\ref{mbt2}(b) also show that the value of $m_{\beta \beta}$ decreases with the increase of the mass of pseudo-Dirac sterile neutrinos and tends to go out of the bound when the pseudo-Dirac sterile neutrinos mass decreases significantly. However, all the cLFV decay can be expressed in GeV range pseudo-Dirac sterile neutrinos mass.

 \section{Summary and Conclusion} 
 We have presented a lepton mass model based on $A_{5}\times Z_{5}\times Z_{9}\times Z_{5}$ symmetry group. This model can give deviation from the GR1 mixing pattern and explain the tiny masses of neutrinos through the linear seesaw mechanism. The charged lepton sector of the model predicts a specific pattern of a mass matrix that can perturb the GR1 mixing pattern, resulting in a unique PMNS matrix $U$ that accurately explains current leptonic mixing data. The specific pattern of $m_{\nu}$ in our model predicts NH as the mass pattern of neutrino masses. With this model, we can also express the leptonic mixing angles $\theta_{12}$, $\theta_{23}$, and $\delta$ in terms of the model parameter $\psi$ and the mixing angle $\theta_{13}$. It predicts a lower octant region for $\theta_{23}$ and $\theta_{12}$ values within the 3$\sigma$ range of current data for the NH case. It can also explain the neutrino masses, and the mass square differences for neutrinos in their current experimental bounds. Again, the model can explain $Br(\mu \rightarrow e\gamma)$ and $Br(\tau \rightarrow e\gamma)$ in the current and projected future sensitivity, and the current sensitivity of $m_{\beta \beta}$ when the mass of quasi-Dirac heavy sterile neutrinos are in the TeV range.  The current sensitivity of the  $Br(\tau \rightarrow \tau\gamma)$ and the current sensitivity of $m_{\beta \beta}$ cannot be reproduced simultaneously by this model. However, it can accommodate the future sensitivity of  $Br(\tau \rightarrow \tau\gamma)$ and the current sensitivity of $m_{\beta \beta}$ when the he mass of quasi-Dirac heavy sterile neutrinos are in the TeV range.  These scenarios give a heavy constraint on neutrino mass parameters, especially on $m$ and $M$, which in turn controls neutrino mass square differences, $\sum m_{i}$, mixing matrix $K_{H}$ and $m_{\beta \beta}$. Like most seesaw cases, mixing matrix $K_{L} \simeq U$ still holds. Moreover, the contribution from the charged lepton mass matrix also gives a significant contribution to the structure of $K_{H}$. Further, we find that in the absence of heavy neutrino contribution, the value of  $m_{\beta \beta}$ remains small and similar for the cases where either  $Br(\mu \rightarrow e\gamma)$ or $Br(\tau \rightarrow e\gamma)$  or $Br(\tau \rightarrow \mu\gamma)$ sensitivity are considered. We observed that $m_{\beta \beta}$ varies significantly when the contribution of the heavy neutrinos masses are introduced in all the cLFV processes. Additionally, we find that the mass of quasi-Dirac heavy sterile neutrinos mass plays a crucial role in explaining current sensitivity of $m_{\beta \beta}$ as there is a non-standard increase in $m_{\beta \beta}$ when the lightest quasi-Dirac heavy sterile neutrino mass decrease from $\sim ~(1.7\times 10^{2}$ TeV, $4\times 10^{2})$ TeV for the case where  $Br(\mu \rightarrow e \gamma)$ is considered. Similar non-standard increase in $m_{\beta \beta}$ is observed when the lightest quasi-Dirac heavy sterile neutrino mass decreases from $\sim~ (5.56 \times 10^{3}$,  $14.08 \times 10^{3})$ TeV in the case where the sensitivity of $Br(\tau \rightarrow e \gamma)$ are considered. On the other hand , the value of  $m_{\beta \beta}$ is relatively lower in $Br(\tau \rightarrow \mu \gamma)$ decay cases but the $m_{\beta \beta}$  follow the similar non-standard increase if the masses of the lightest quasi-Dirac heavy sterile neutrino mass decreases from  $\sim (4.92-9.90 )10^{3}$ TeV range.

  \section*{Acknowledgements}

One of us (VP)  wishes to thank  Department of Science and Technology (DST), Government of India for providing INSPIRE Fellowship. We are thankful to Dr. Happy Borgohain, Department of Physics, Indian Institute for Technology, Guwahati , for fruitful discussions.

\appendix  
\section{$A_{5}$ Group}
$A_{5}$ is the even permutation group of 5 objects with $\frac{5!}{2}$ elements. It has five irreducible representations, namely {\bf{1}}, {\bf{3}}, ${\bf{3^{'}}}$, {\bf{4}} and {\bf{5}}. All the elements of the group can be generated by two elements S and T. The generators S and T satisfy the relation,
Two generators $S$ and $T$ obey the relation
\begin{equation}
S^{2}=T^{5}=(ST)^3=1
\end{equation} 

To show the tensor product rules of the $A_{5}$ discrete group, we assign $a=(a_{1}, a_{2},a_{3})^{T}$ and $b=(b_{1}, b_{2}. b_{3})^{T}$ to the {\bf{3}} representation, while $a^{'}=(a^{'}_{1}, a^{'}_{2}, a^{'}_{3})^{T}$ and $b^{'}=(b^{'}_{1}, b^{'}_{2}, b^{'}_{3})^{T}$ belong to the {$\bf{3^{'}}$} representation. $c=(c_{1},c_{2},c_{3},c_{4}c_{5})^{T}$ and $d=d_{1}, d_{2}, d_{3}, d_{4}, d_{5})^{T}$ are pentaplets; $f=(f_{1}, f_{2}, f_{3}, f_{4})^{T}$ and $g=(g_{1}, g_{2}, g_{3}, g_{4})^{T}$ are tetraplets.\\

${\bf{3}}\otimes{\bf{3}}={\bf{3}}_{a}+({\bf{1}}+{\bf{5}})_{s}$\\
$~~~~~~~~~~~~~{\bf{1}}=a_{1}b_{1}+a_{2}b_{3}+a_{3}b_{2}$\\
$~~~~~~~~~~~~~{\bf{3}}=(a_{2}b_{3}-a_{3}b_{2}, a_{1}b_{2}-a_{2}b_{1},a_{3}b_{1}-a_{1}b_{3})^{T}$\\
$~~~~~~~~~~~~~{\bf{5}}=(a_{1}b_{1}-\frac{a_{2}b_{3}}{2}-\frac{a_{3}b_{2}}{2}, \frac{\sqrt{3}}{2}(a_{1}b_{2}+a_{2}b_{1}),-\sqrt{\frac{3}{2}}a_{2}b_{2}, -\sqrt{\frac{3}{2}}a_{3}b_{3}, -\frac{\sqrt{3}}{2}(a_{1}b_{3}+a_{3}b_{1}))^{T}$\\

${\bf{3^{'}}}\otimes {\bf{3^{'}}}={\bf{3^{'}}}_{a}+({\bf{1}}+{\bf{5}})_{s}$\\
$~~~~~~~~~~~~~{\bf{1}}=a_{1}^{'}b_{1}^{'}+a_{2}^{'}b_{3}^{'}+a_{3}^{'}b_{2}^{'}$\\
$~~~~~~~~~~~~~{\bf{3^{'}}}=(a_{2}^{'}b_{3}^{'}-a_{3}^{'}b_{2}^{'}, a_{1}^{'}b_{2}^{'}-a_{2}^{'}b_{1}^{'},a_{3}^{'}b_{1}^{'}-a_{1}^{'}b_{3}^{'})^{T}$\\
$~~~~~~~~~~~~~{\bf{5}}=(a_{1}^{'}b_{1}^{'}-\frac{a_{2}^{'}b_{3}^{'}}{2}-\frac{a_{3}^{'}b_{2}^{'}}{2}, \frac{\sqrt{3}}{2}(a_{1}^{'}b_{2}^{'}+a_{2}^{'}b_{1}^{'}),-\sqrt{\frac{3}{2}}a_{2}^{'}b_{2}^{'}, -\sqrt{\frac{3}{2}}a_{3}^{'}b_{3}^{'}, -\frac{\sqrt{3}}{2}(a_{1}^{'}b_{3}^{'}+a_{3}^{'}b_{1}^{'}))^{T}$\\

${\bf{3}}\otimes {\bf{4}}={\bf{3^{'}}}+{\bf{4}}+{\bf{5}}$\\
$~~~~~~~~~~~~~{\bf{3^{'}}}=(a_{2}g_{4}-a_{3}g_{1},\frac{1}{\sqrt{2}}(\sqrt{2} a_{1}g_{2}+a_{2}g_{1}+a_{3}g_{3})-\frac{1}{\sqrt{2}}(\sqrt{2}a_{1}g_{3}+a_{2}g_{2}+a_{3}g_{4}))^{T}$\\
$~~~~~~~~~~~~~{\bf{4}}=(a_{1}g_{1}+\sqrt{2}a_{3}g_{2},-a_{1}g_{2}+\sqrt{2}a_{2}g_{1},a_{1}g_{3}-\sqrt{2}a_{3}g_{4},-a_{1}g_{4}-\sqrt{2}a_{2}g_{3})^{T}$\\
$
~~~~~~~~~~~~~{\bf{5}}=(a_{3}g_{1}+a_{2}g_{4},\sqrt{\frac{2}{3}}(\sqrt{2}a_{1}g_{1}-a_{3}g_{2}),\frac{1}{\sqrt{6}}(\sqrt{2}a_{1}g_{2}-3a_{3}g_{3}+a_{2}g_{1}),\frac{1}{\sqrt{6}}(\sqrt{2}a_{1}g_{3}-\\
~~~~~~~~~~~~~~~~~~~3a_{2}g_{2}+a_{3}g_{4}),\sqrt{\frac{2}{3}}(-\sqrt{2}a_{1}g_{4}+a_{2}g_{3}))^{T}$\\

${\bf{3^{'}}}\otimes {\bf{4}}={\bf{3}}+{\bf{4}}+{\bf{5}}$\\

$~~~~~~~{\bf{3}}=(a_{2}^{'}g_{3}-a_{3}^{'}g_{2},\frac{1}{\sqrt{2}}(\sqrt{2}a_{1}^{'}g_{1}+a_{2}^{'}g_{4}-a_{3}^{'}g_{3}), \frac{1}{\sqrt{2}}(-\sqrt{2}a_{1}^{'}g_{4}+a_{2}^{'}g_{2}-a_{3}^{'}g_{1}))^{T}$\\
$~~~~~~~~~~~~{\bf{4}}=(a_{1}^{'}g_{1}+\sqrt{2}a_{3}^{'}g_{3},~ a_{1}^{'}g_{2}-\sqrt{2}a_{3}^{'}g_{4},~ -a_{1}^{'}g_{3}+\sqrt{2} a_{2}^{'}g_{1},~-a_{1}^{'}g_{4}-\sqrt{2}a_{2}^{'}g_{2})^{T}$\\
$~~~~~~~~~~~~~{\bf{5}}=(a_{3}^{'}+g_{2}+a_{2}^{'}g_{3},\frac{1}{\sqrt{6}}(\sqrt{2}a_{1}^{'}g_{1}-3a_{2}^{'}g_{4}-a_{3}^{'}g_{3}),\sqrt{\frac{2}{3}}(\sqrt{2}a_{1}^{'}g_{2}+a_{3}^{'}g_{4}),-\sqrt{\frac{2}{3}}(\sqrt{2}a_{1}^{'}g_{3}+\\
~~~~~~~~~~~~~~~~~~~~~~a_{2}^{'}g_{1}),\frac{1}{\sqrt{6}}(-\sqrt{2}a_{1}^{'}g_{4}+3a_{3}^{'}g_{1}+a_{2}^{'}g_{2}))^{T}$\\

$3\otimes 5=3+3^{'}+4+5$\\

$~~~~~~~~3=(\frac{2 a_{1}c_{1}}{\sqrt{3}}+a_{3}c_{2}-a_{2}c_{5},-\frac{a_{2}c_{1}}{\sqrt{3}}+a_{1}c_{2}-\sqrt{2}a_{3}c_{3},-\frac{a_{3}c_{1}}{\sqrt{3}}-a_{1}c_{5}-\sqrt{2}a_{2}c_{4})^{T}$\\
$~~~~~~~~~~~~~3^{'}=(a_{1}c_{1}+\frac{a_{2}c_{5}-a_{3}c_{2}}{\sqrt{3}}, \frac{a_{1}c_{3}+\sqrt{2}(a_{3}c_{4}-a_{2}c_{2})}{\sqrt{3}}, \frac{a_{1}c_{4}+\sqrt{2}(a_{2}c_{3}+a_{3}c_{5})}{\sqrt{3}})^{T}$\\
$~~~~~~~~~~~~~~~~~~4=(4a_{1}c_{2}+2\sqrt{3}a_{2}c_{1}+\sqrt{2}a_{3}c_{3},2a_{1}c_{3}-2\sqrt{2}a_{2}c_{2}-3\sqrt{2}a_{3}c_{4}, 2a_{1}c_{4}-3\sqrt{2}a_{2}c_{3}+2\sqrt{2}a+{3}\\ ~~~~~~~~~~~~~~~~~~c_{5},-4a_{1}c_{5}+\sqrt{2}a_{2}c_{4}+2\sqrt{3}a_{3}c_{1})^{T}$\\
$~~~~~~~~~~~~~5=(a_{2}c_{5}+a_{3}c_{2},~ a_{2}c_{1}-\frac{a_{1}c_{1}+\sqrt{2}a_{3}c_{3}}{\sqrt{3}},-\frac{2a_{1}c_{3}+\sqrt{2}a_{2}c_{2}}{\sqrt{3}}, \frac{2a_{1}c_{4}-\sqrt{2}a_{3}c_{5}}{\sqrt{3}},~ a_{3}c_{1}+\frac{a_{1}c_{5}-\sqrt{2}a_{2}c_{4}}{\sqrt{3}})^{T}$\\

$3^{'}\otimes 5=3+3^{'}+4+5$\\
$~~~~~~~~~~~~~3=(a_{1}^{'}c_{1}+\frac{a_{3}^{'}c_{3}+a_{2}^{'}c_{4}}{\sqrt{3}}, \frac{-a_{1}^{'}c_{2}+\sqrt{2}(a_{3}^{'}c_{4}+a_{2}^{'}c_{5})}{\sqrt{3}}, \frac{a_{1}^{'}c_{5}+\sqrt{2}(a_{2}^{'}c_{3}-a_{3}^{'}c_{2})}{\sqrt{3}})^{T}$\\
$~~~~~~~~~~~~~3^{'}=(\frac{2a_{1}^{'}}{\sqrt{3}}-a_{3}^{'}c_{3}-a_{2}^{'}c_{4},-\frac{a_{2}^{'}c_{1}}{\sqrt{3}}-a_{1}^{'}c_{3}-\sqrt{2}a_{2}^{'}c_{5},-\frac{a_{3}^{'}c_{1}}{\sqrt{3}}-a_{1}^{'}c_{4}+\sqrt{2}a_{2}^{'}c_{2})^{T}$\\
$~~~~~~~~~~~~~4=(2a_{1}^{'}c_{2}+3\sqrt{2}a_{2}^{'}c_{5}-2\sqrt{2}a_{3}^{'}c_{4},-4a_{1}^{'}c_{3}+2\sqrt{3}a_{2}^{'}c_{1}+\sqrt{2}a_{3}^{'}c_{5}, -4a_{1}^{'}c_{4}-\sqrt{2}a_{2}^{'}c_{2}+\\~~~~~~~~~~~~~~~~~~2\sqrt{3}a_{3}^{'}c_{1},-2a_{1}^{'}c_{5}-2\sqrt{2}a_{2}^{'}c_{3}-3\sqrt{2}a_{3}^{'}c_{2})^{T}$\\
$~~~~~~~~~~~~~~5=(a_{2}^{'}c_{4}-a_{3}^{'}c_{3}, \frac{2a_{1}^{'}c_{2}+\sqrt{2}a_{3}^{'}c_{4}}{\sqrt{3}},-a_{2}^{'}c_{1}-\frac{a_{1}^{'}c_{3}-\sqrt{2}a_{3}^{'}c_{5}}{\sqrt{3}}, a_{3}^{'}c_{1}+\frac{a_{1}^{'}c_{4}+\sqrt{2}a_{2}^{'}c_{2}}{\sqrt{3}}, \frac{-2a_{1}^{'}c_{5}+\sqrt{2}a_{2}^{'}c_{3}}{\sqrt{3}})^{T}$\\

$4\otimes 4=(3+3^{'})_{a}+(1+4+5)_{s}$\\
$~~~~~~~~~~~~~1=f_{1}g_{4}+f_{2}g_{	3}+f_{3}g_{2}+f_{4}g_{1}$\\
$~~~~~~~~~~~~~3=f_{1}g_{4}-f_{4}g_{1}+f_{3}g_{2}-f_{2}g_{3},\sqrt{2}(f_{2}g_{4}-f_{4}g_{2}), \sqrt{2}(f_{1}g_{3}-f_{3}g_{1}))^{T}$\\
$~~~~~~~~~~~~~3^{'}=f_{1}g_{4}-f_{4}g_{1}+f_{2}g_{3}-f_{3}g_{2}, \sqrt{2}(f_{3}g_{4}-f_{4}g_{3}), \sqrt{2}(f_{1}g_{2}-f_{2}g_{1}))^{T}$\\
$~~~~~~~~~~~~~4=(f_{3}g_{3}-f_{4}g_{2}, f_{1}g_{1}+f_{3}g_{4}+f_{4}g_{3}, -f_{4}g_{4}-f_{1}g_{2}-f_{2}g_{1},-f_{2}g_{2}+f_{1}g_{3}+f_{3}g_{1})^{T}$\\
$~~~~~~~~~~~~~5=(f_{1}g_{4}+f_{4}g_{1}-f_{3}g_{2}-f_{2}g_{3},-\sqrt{\frac{2}{3}}(2f_{3}g_{3}+f_{2}g_{4}+f_{4}g_{2}), \sqrt{\frac{2}{3}}(-2f_{1}g_{1}+f_{3}g_{4}+\\
~~~~~~~~~~~~~~~~~~~f_{4}g_{3}), \sqrt{\frac{2}{3}}(-2f_{4}g_{4}+f_{2}g_{1}+f_{1}g_{2}),\sqrt{\frac{2}{3}}(2f_{2}g_{2}+f_{1}g_{3}+f_{3}g_{1}))^{T}$\\

$4\otimes 5=3+3^{'}+4+5+5$\\
$~~~~~~~~~~~~~3=4f_{1}c_{5}-4f_{4}c_{2}-2f_{2}c_{3}-2f_{2}c_{4},-2\sqrt{3}f_{1}c_{1}-\sqrt{2}(2f_{2}c_{5}-3f_{3}c_{4}+f_{4}c_{3}), \sqrt{2}(-f_{4}c_{4}+\\
~~~~~~~~~~~~~~~~~3f_{2}c_{3}+2f_{3}c_{2}-2\sqrt{3}f_{4}c_{1})^{T}$\\
$~~~~~~~~~~~~~3^{'}=(2f_{1}c_{5}-2f_{4}c_{2}+4f_{3}c_{3}+4f_{2}c_{4}, -2\sqrt{3}f_{2}c_{1}+\sqrt{2}(2f_{4}c_{4}+3f_{1}c_{2}-f_{3}c_{5}),\sqrt{2}(f_{2}c_{2}-\\
~~~~~~~~~~~~~~~~~~~3f_{4}c_{5}+2f_{1}c_{3})-2\sqrt{3}f_{3}c_{1})^{T}$\\
$~~~~~~~~~~~~4=(3f_{1}c_{1}+\sqrt{6}(f_{2}c_{5}+f_{5}c_{4}-2f_{4}c_{3}), -3f_{2}c_{1}+\sqrt{6}(f_{4}c_{4}-f_{1}c_{2}+2f_{3}c_{5}), -3f_{3}c_{1}+\\~~~~~~~~~~~~~~~~~~\sqrt{6}(f_{1}c_{3}+f_{4}c_{5}-2f_{2}c_{2}), 3f_{4}c_{1}+\sqrt{6}(f_{2}c_{3}-f_{3}c_{2}-2f_{1}c_{4}))^{T}$\\
$~~~~~~~~~~~~5_{1}=(f_{1}c_{5}+2f_{2}c_{4}-2f_{3}c_{3}+f_{4}c_{2}, -2f_{1}c_{1}+\sqrt{6}f_{2}c_{5}, f_{2}c_{1}+\sqrt{\frac{3}{2}}(-f_{1}c_{2}-f_{3}c_{5}+\\
~~~~~~~~~~~~~~~~~~~2f_{4}c_{4}),-f_{3}c_{1}-\sqrt{\frac{3}{2}}(f_{2}c_{2}+f_{4}c_{5}+
2f_{1}c_{3}), -2f_{4}c_{1}-\sqrt{6}f_{3}c_{2})^{T}$\\
$~~~~~~~~~~~~5_{2}=(f_{2}c_{4}-f_{3}c_{3}, -f_{1}c_{1}+frac{2f_{1}c_{5}-f_{3}c_{4}-f_{4}c_{3})}{\sqrt{5}}, -\sqrt{\frac{2}{3}}(f_{1}c_{1}+f_{3}c_{5}-\\
~~~~~~~~~~~~~~~~~~f_{4}c_{4}), -\sqrt{\frac{2}{3}}(f_{1}c_{3}+f_{2}c_{2}+f_{4}c_{5}), -f_{4}c_{1}-\frac{2f_{3}c_{2}+f_{1}c_{4}+f_{2}c_{3}}{\sqrt{6}})^{T}$\\

$5\otimes 5=(3+3^{'}+4)_{a}+(1+4+5+5)_{s}$\\
$~~~~~~~~~~~~~3=(2(c_{4}d_{3}-c_{3}d_{4})+c_{2}d_{5}-c_{5}d_{2},\sqrt{3}(c_{2}d_{1}-c_{1}d_{2})+\sqrt{2}(c_{3}d_{5}-c_{5}d_{3}), \sqrt{3}(c_{5}d_{1}-\\
~~~~~~~~~~~~~~~~~~~c_{1}d_{5})+\sqrt{2}(c_{4}d_{2}-c_{2}d_{4}))^{T}$\\
$~~~~~~~~~~~~~3^{'}=(2(c_{2}d_{5}-c_{5}d_{2})+c_{3}d_{4}-c_{4}d_{2}, \sqrt{3}(c_{3}d_{1}-c_{1}d_{3})+\sqrt{2}(c_{4}d_{5}-c_{5}d_{4}), \sqrt{3}(c_{1}d_{4}-\\
~~~~~~~~~~~~~~~~~~~~c_{4}d_{1})+\sqrt{2}(c_{3}d_{2}-c_{2}d_{3}))^{T}$\\
$~~~~~~~~~~~~~4_{s}=((c_{1}d_{2}+c_{2}d_{1})-\frac{c_{3}d_{5}+c_{5}d_{3})-4c_{4}d_{4}}{\sqrt{6}}, -(c_{1}d_{3}+c_{3}d_{1})-\frac{(c_{4}d_{5}+c_{5}d_{4})-4c_{2}d_{2}}{\sqrt{6}}, -(c_{1}d_{4}+\\
~~~~~~~~~~~~~~~~~~~c_{4}d_{1})-\frac{(c_{2}d_{3}+c_{3}d_{2})+4c_{5}d_{5}}{\sqrt{6}}, (c_{1}d_{5}+c_{5}d_{1})-\frac{(c_{2}d_{4}+c_{4}d_{2}+4c_{3}d_{3}}{\sqrt{6}})^{T}$\\
$~~~~~~~~~~~~~4_{a}=((c_{1}d_{2}-c_{2}d_{1})+\sqrt{\frac{3}{2}}(c_{3}d_{5}-c_{5}d_{3}), (c_{1}d_{3}-c_{3}d_{1})+\sqrt{\frac{3}{2}}(c_{4}d_{5}-c_{5}d_{4}), (c_{4}d_{1}-\\~~~~~~~~~~~~~~~~~~~~
c_{1}d_{4})+\sqrt{\frac{3}{2}}(c_{3}d_{2}-c_{2}d_{3}), (c_{1}d_{5}-c_{5}d_{1})+\sqrt{\frac{3}{2}}(c_{4}d_{2}-c_{2}d_{4}))^{T}$\\
$~~~~~~~~~~~~~5_{1}=(c_{1}d_{1}+c_{2}d_{5}+c_{5}d_{2}+\frac{c_{3}d_{4}+c_{4}d_{3}}{2}, -(c_{1}d_{2}+c_{2}d_{1})+\sqrt{\frac{3}{2}}c_{4}d_{4}, \frac{1}{2}(c_{1}d_{3}+c_{3}d_{1}-\\
~~~~~~~~~~~~~~~~~~~~~~\sqrt{6}(c_{4}d_{5}+c_{5}d_{4})),\frac{1}{2}(c_{1}d_{4}+c_{4}d_{1}+\sqrt{6}(c_{2}d_{3}+c_{3}d_{2})), (c_{1}d_{5}+c_{5}d_{1})-\sqrt{\frac{3}{2}}c_{3}d_{3})^{T}$\\
$~~~~~~~~~~~~~~~~~~~5_{2}=(\frac{2c_{1}d_{1}+c_{2}d_{5}+c_{5}d_{2}}{2}, \frac{-3(c_{1}d_{2}+c_{2}d_{1})+\sqrt{6}(2c_{4}d_{4}+c_{3}d_{5}+c_{5}d_{3})}{6}, \frac{-2c_{4}d_{5}+2c_{5}d_{4}+c_{2}d_{2}}{\sqrt{6}}, \frac{2c_{2}d_{3}+3c_{3}d_{2}-c_{5}d_{5}}{\sqrt{6}},\\~~~~~~~~~~~~~~~~~~~~
 -\frac{3(c_{1}d_{5}+c_{5}d_{1})+\sqrt{6}(-2c_{3}d_{3}+c_{2}d_{4}+c_{4}d_{2})}{6})^{T}$\\
$~~~~~~~~~~~~~~1=c_{1}d_{1}+c_{3}d_{4}+c_{4}d_{3}-c_{2}d_{5}-c_{5}d_{2}$  
\section{Vacuum Alignment}
As mentioned in Section II, the VEV of the triplets and the pentaplets  of the model can be obtained from the superpotential in Eq .(\ref{w1}) 
\begin{align}
W=&M_{o}\eta_{1} \eta_{2}+ M_{1} \phi_{1}^{2}+M_{2} (\phi_{2}\phi_{3})+g(\phi_{1}\phi_{2}\phi_{3})+g_{1}\eta_{1}(\phi_{5}^{2})+g_{2}(\phi_{5}^{3})_{1}\nonumber\\ 
&+g_{3}(\phi_{5}^{3})_{2}+\frac{g_{4}}{3}\eta_{1}^3+\frac{g_{5}}{3}\eta^{3}_{2}+...
\end{align}

Here, the desired VEV alignments are obtained without using driving fields and the minima are derived by analysing the F-terms of the flavons themselves. The VEV of the  triplet flavons can be determined by analysing the minima  given below:
\begin{equation}
\frac{\partial W}{\partial \phi_{1_{1}}}=2M_{1}\phi_{1_{1}}+g(\phi_{2_{2}}\phi_{3_{3}}-\phi_{2_{3}}\phi_{3_{2}})=0
\end{equation}
\begin{equation}
\frac{\partial W}{\partial \phi_{1_{2}}}=2M_{1}\phi_{1_{3}}+g(\phi_{2_{3}}\phi_{3_{1}}-\phi_{2_{1}}\phi_{3_{3}})=0
\end{equation}
\begin{equation}
\frac{\partial W}{\partial \phi_{1_{3}}}=2M_{1}\phi_{1_{2}}+g(\phi_{2_{1}}\phi_{3_{2}}-\phi_{2_{2}}\phi_{3_{1}})=0
\end{equation}
\begin{equation}
\frac{\partial W}{\partial \phi_{2_{1}}}=M_{2}\phi_{3_{1}}+g(\phi_{1_{3}}\phi_{3_{2}}-\phi_{1_{2}}\phi_{3_{3}})=0
\end{equation}
\begin{equation}
\frac{\partial W}{\partial \phi_{2_{2}}}=M_{2}\phi_{3_{3}}+g(\phi_{1_{1}}\phi_{3_{3}}-\phi_{1_{3}}\phi_{3_{1}})=0
\end{equation}
\begin{equation}
\frac{\partial W}{\partial \phi_{2_{3}}}=M_{2}\phi_{3_{2}}+g(\phi_{1_{2}}\phi_{3_{1}}-\phi_{1_{1}}\phi_{3_{2}})=0
\end{equation}
\begin{equation}
\frac{\partial W}{\partial \phi_{3_{1}}}=M_{2}\phi_{2_{1}}+g(\phi_{1_{2}}\phi_{2_{3}}-\phi_{1_{3}}\phi_{2_{2}})=0
\end{equation}
\begin{equation}
\frac{\partial W}{\partial \phi_{3_{2}}}=M_{2}\phi_{2_{3}}+g(\phi_{1_{3}}\phi_{2_{1}}-\phi_{1_{1}}\phi_{2_{3}})=0
\end{equation}
\begin{equation}
\frac{\partial W}{\partial \phi_{3_{3}}}=M_{2}\phi_{2_{2}}+g(\phi_{1_{1}}\phi_{2_{2}}-\phi_{1_{2}}\phi_{2_{1}})=0
\end{equation}

Following the same procedure and consideration as the Ref. \cite{Feruglio:2011qq}, one can determine the solutions for the above equations. One of such solution is given by

\begin{align}
<\phi_{1}>&=-\frac{M_2}{g}(1,0,0)\nonumber \\
<\phi_{2}>&=-\frac{\sqrt{2M_{1}M_{2}}}{g}(0,0,1)\nonumber\\
<\phi_{3}>&=-\frac{\sqrt{2M_{1}M_{2}}}{g}(0,1,0)
\end{align}

The other possible solutions are shown in Ref.  \cite{Feruglio:2011qq}. The VEV of the pentaplets can also be determined. Here, one need to discuss sets of minima

\begin{equation}
\frac{\partial W}{\partial \eta_{2}}=0
\label{et}
\end{equation} 
\begin{equation}
\frac{\partial W}{\partial \eta_{1}}=0
\end{equation} 
\begin{equation}
\frac{\partial W}{\partial \phi_{5}}=0
\label{p5a}
\end{equation} 

The Eq.(\ref{et}) gives
\begin{equation}
\eta_{1}=-\frac{g_{5}}{M_{1}}\eta_{2}^{2}
\end{equation}

The  equations  given by Eq. (\ref{p5a})  are conveniently solved using Cummins-Patera basis \cite{Cummins:1988dr,Luhn:2008ey} for the generators S and T in Ref \cite{Feruglio:2011qq}. Following the same procedure, one can write the components of $\phi_{5}$ in Cummins-Patera basis as 

\begin{equation}
\phi_{5}=(x_{1},x_{2},x_{3}, z, \bar{z}), 
\end{equation}

 where $x_{i}(i=1,2,3)$, $z$ and $\bar{z}$ are treated as independent complex quantities. Again, in this basis, superpotential $W$  terms containing $\phi_{5}$ can be written as
 
\begin{align}
W=&g_{1}\eta_{1}(x_{1}^{2}+x_{2}^{2}+x_{3}^{2}+2z\bar{z})+g_{2}(z^{3}-\bar{z}^{3}-3(x_{1}^{2}+\omega^{2}x_{2}^{2}+\omega x_{3}^{2})z+3(x_{1}^{2}+\omega x_{2}^{2}+\omega^{2} x_{3}^{2})\bar{z})+\nonumber\\
&g_{3}(z^{3}+\bar{z}^{3}+(x_{1}^{2}+\omega^{2}x_{2}^{2}+\omega x_{3}^{2})z+(x_{1}^{2}+\omega x_{2}^{2}+\omega^{2} x_{3}^{2}-4x_{1}x_{2}x_{3})\bar{2})+...
\end{align}
where $\omega=e^{\frac{2\pi i}{3}}$. By using the set of minima mentioned above, one of the solutions obtained is given by
 \begin{equation}
 x_{1}=x_{2}=x_{3}=0
 \label{x1}
\end{equation}  
\begin{equation}
z=-\frac{2g_{1}g_{5}}{3M_{1}(g_{2}-g_{3})^{1/3}(g_{2}+g_{3})^{2/3}}
\label{x2}
\end{equation}
\begin{equation}
\bar{z}=\frac{2g_{1}g_{5}}{3M_{1}(g_{2}-g_{3})^{2/3}(g_{2}+g_{3})^{1/3}}
\label{x3}
\end{equation}

After making same consideration as Ref \cite{Feruglio:2011qq} and by relating the Cummins-Patera basis to the basis used in the model with the help of unitary transformation, one can translates the minimum obtain in the Eqs. (\ref{x1}-\ref{x3}) to the basis utilised in our model as
\begin{equation}
<\phi_{5}>=(-\sqrt{\frac{2}{3}}(p+q),-p,q,q,p)v_{s}
\end{equation}

Here, the VEV of the pentaplet and the triplets are well established . The detail calculation can be found in Ref \cite{Feruglio:2011qq}. Further, in the model, we also introduced singlet scalars $\xi_{1}$, $\xi_{2}$ and $\xi_{3}$ and they  are simply considered to have the VEV $u_{1}$, $u_{2}$  and $u_{3}$, respectively as in Ref \cite{Hernndez2023}. 
\bibliographystyle{ieeetr}
\bibliography{vreff1}

\begin{thebibliography}{10}

\bibitem{KamLAND:2002uet}
K.~Eguchi {\em et~al.}, ``{First results from KamLAND: Evidence for reactor
  anti-neutrino disappearance},'' {\em Phys. Rev. Lett.}, vol.~90, p.~021802,
  2003.

\bibitem{SNO:2002tuh}
Q.~R. Ahmad {\em et~al.}, ``{Direct evidence for neutrino flavor transformation
  from neutral current interactions in the Sudbury Neutrino Observatory},''
  {\em Phys. Rev. Lett.}, vol.~89, p.~011301, 2002.

\bibitem{Super-Kamiokande:1998kpq}
Y.~Fukuda {\em et~al.}, ``{Evidence for oscillation of atmospheric
  neutrinos},'' {\em Phys. Rev. Lett.}, vol.~81, pp.~1562--1567, 1998.

\bibitem{DoubleChooz:2011ymz}
Y.~Abe {\em et~al.}, ``{Indication of Reactor $\bar{\nu}_e$ Disappearance in
  the Double Chooz Experiment},'' {\em Phys. Rev. Lett.}, vol.~108, p.~131801,
  2012.

\bibitem{Lasserre:2012ax}
T.~Lasserre, G.~Mention, M.~Cribier, A.~Collin, V.~Durand, V.~Fischer,
  J.~Gaffiot, D.~Lhuillier, A.~Letourneau, and M.~Vivier, ``{Comment on Phys.
  Rev. Lett. 108, 191802 (2012): 'Observation of Reactor Electron Antineutrino
  Disappearance in the RENO Experiment'},'' 5 2012.

\bibitem{Ling:2013fta}
J.~Ling, ``{Observation of electron-antineutrino disappearance at Daya Bay},''
  {\em AIP Conf. Proc.}, vol.~1560, no.~1, pp.~206--210, 2013.

\bibitem{McDonald:2016ixn}
A.~B. McDonald, ``{Nobel Lecture: The Sudbury Neutrino Observatory: Observation
  of flavor change for solar neutrinos},'' {\em Rev. Mod. Phys.}, vol.~88,
  no.~3, p.~030502, 2016.

\bibitem{GERDA:2019ivs}
M.~Agostini {\em et~al.}, ``{Probing Majorana neutrinos with double-$\beta$
  decay},'' {\em Science}, vol.~365, p.~1445, 2019.

\bibitem{CUORE:2019yfd}
D.~Q. Adams {\em et~al.}, ``{Improved Limit on Neutrinoless Double-Beta Decay
  in $^{130}$Te with CUORE},'' {\em Phys. Rev. Lett.}, vol.~124, no.~12,
  p.~122501, 2020.

\bibitem{KamLAND-Zen:2016pfg}
A.~Gando {\em et~al.}, ``{Search for Majorana Neutrinos near the Inverted Mass
  Hierarchy Region with KamLAND-Zen},'' {\em Phys. Rev. Lett.}, vol.~117,
  no.~8, p.~082503, 2016.
\newblock [Addendum: Phys.Rev.Lett. 117, 109903 (2016)].

\bibitem{KATRIN:2019yun}
M.~Aker {\em et~al.}, ``{Improved Upper Limit on the Neutrino Mass from a
  Direct Kinematic Method by KATRIN},'' {\em Phys. Rev. Lett.}, vol.~123,
  no.~22, p.~221802, 2019.

\bibitem{Planck:2018nkj}
N.~Aghanim {\em et~al.}, ``{Planck 2018 results. I. Overview and the
  cosmological legacy of Planck},'' {\em Astron. Astrophys.}, vol.~641, p.~A1,
  2020.

\bibitem{deSalas:2020pgw}
P.~F. de~Salas, D.~V. Forero, S.~Gariazzo, P.~Mart\'\i{}nez-Mirav\'e, O.~Mena,
  C.~A. Ternes, M.~T\'ortola, and J.~W.~F. Valle, ``{2020 global reassessment
  of the neutrino oscillation picture},'' {\em JHEP}, vol.~02, p.~071, 2021.

\bibitem{ch11ch11}
R.~Bernstein and P.~S. Cooper, ``Charged lepton flavor violation: An
  experimenter’s guide,'' {\em Physics Reports}, vol.~532, no.~2, pp.~27--64,
  2013.
\newblock Charged Lepton Flavor Violation: An Experimenter's Guide.

\bibitem{annurev-nucl-102912-144530}
S.~Mihara, J.~Miller, P.~Paradisi, and G.~Piredda, ``Charged lepton
  flavor–violation experiments,'' {\em Annual Review of Nuclear and Particle
  Science}, vol.~63, no.~1, pp.~531--552, 2013.

\bibitem{Ding:2011cm}
G.-J. Ding, L.~L. Everett, and A.~J. Stuart, ``{Golden Ratio Neutrino Mixing
  and $A_5$ Flavor Symmetry},'' {\em Nucl. Phys. B}, vol.~857, pp.~219--253,
  2012.

\bibitem{Feruglio:2011qq}
F.~Feruglio and A.~Paris, ``{The Golden Ratio Prediction for the Solar Angle
  from a Natural Model with A5 Flavour Symmetry},'' {\em JHEP}, vol.~03,
  p.~101, 2011.

\bibitem{Everett:2008et}
L.~L. Everett and A.~J. Stuart, ``{Icosahedral (A(5)) Family Symmetry and the
  Golden Ratio Prediction for Solar Neutrino Mixing},'' {\em Phys. Rev. D},
  vol.~79, p.~085005, 2009.

\bibitem{CarcamoHernandez:2022bka}
A.~E. C\'arcamo~Hern\'andez and I.~de~Medeiros~Varzielas, ``{An A5 inverse
  seesaw model with perturbed golden ratio mixing},'' {\em Nucl. Phys. B},
  vol.~985, p.~116031, 2022.

\bibitem{Lei:2020nik}
M.~Lei and J.~D. Wells, ``{Minimally modified $A_4$ Altarelli-Feruglio model
  for neutrino masses and mixings and its experimental consequences},'' {\em
  Phys. Rev. D}, vol.~102, no.~1, p.~016023, 2020.

\bibitem{BORAH201959}
D.~Borah and B.~Karmakar, ``Linear seesaw for dirac neutrinos with a4 flavour
  symmetry,'' {\em Physics Letters B}, vol.~789, pp.~59--70, 2019.

\bibitem{Puyam:2022mej}
V.~Puyam, S.~R. Singh, and N.~N. Singh, ``{Deviation from Tribimaximal mixing
  using A4 flavour model with five extra scalars},'' {\em Nucl. Phys. B},
  vol.~983, p.~115932, 2022.

\bibitem{Puyam:2022efv}
V.~Puyam, ``{Nonzero $\theta_{13}$ with $A_4 \times Z_4$ Flavor Symmetry
  Group},'' {\em LHEP}, vol.~2022, p.~340, 2022.

\bibitem{Gupta:2011ct}
S.~Gupta, A.~S. Joshipura, and K.~M. Patel, ``{Minimal extension of
  tri-bimaximal mixing and generalized $Z_2 \to Z_2$ symmetries},'' {\em Phys.
  Rev. D}, vol.~85, p.~031903, 2012.

\bibitem{Gautam:2020bnx}
N.~Gautam, R.~Krishnan, and M.~K. Das, ``{Effect of Sterile Neutrino on
  Low-Energy Processes in Minimal Extended Seesaw With \ensuremath{\Delta}(96)
  Symmetry and TM1 Mixing},'' {\em Front. in Phys.}, vol.~0, p.~417, 2021.

\bibitem{Vien:2015fhk}
V.~V. Vien, H.~N. Long, and D.~P. Khoi, ``{Neutrino Mixing with Non-Zero
  $\theta_{13}$ and CP Violation in the 3-3-1 Model Based on $S_4$ Flavor
  Symmetry},'' {\em Int. J. Mod. Phys. A}, vol.~30, no.~17, p.~1550102, 2015.

\bibitem{CarcamoHernandez:2019eme}
A.~E. C\'arcamo~Hern\'andez and S.~F. King, ``{Littlest Inverse Seesaw
  Model},'' {\em Nucl. Phys. B}, vol.~953, p.~114950, 2020.

\bibitem{Kobayashi:2003fh}
T.~Kobayashi, J.~Kubo, and H.~Terao, ``{Exact S(3) symmetry solving the
  supersymmetric flavor problem},'' {\em Phys. Lett. B}, vol.~568, pp.~83--91,
  2003.

\bibitem{Morisi:2011pm}
S.~Morisi, K.~M. Patel, and E.~Peinado, ``{Model for T2K indication with
  maximal atmospheric angle and tri-maximal solar angle},'' {\em Phys. Rev. D},
  vol.~84, p.~053002, 2011.

\bibitem{Feruglio:2019ybq}
F.~Feruglio and A.~Romanino, ``{Lepton flavor symmetries},'' {\em Rev. Mod.
  Phys.}, vol.~93, no.~1, p.~015007, 2021.

\bibitem{Ding:2023htn}
G.-J. Ding and S.~F. King, ``{Neutrino mass and mixing with modular
  symmetry},'' {\em Rept. Prog. Phys.}, vol.~87, no.~8, p.~084201, 2024.

\bibitem{Ding:2024ozt}
G.-J. Ding and J.~W.~F. Valle, ``{The symmetry approach to quark and lepton
  masses and mixing},'' 2 2024.

\bibitem{RODEJOHANN2009267}
W.~Rodejohann, ``Unified parametrization for quark and lepton mixing angles,''
  {\em Physics Letters B}, vol.~671, no.~2, pp.~267--271, 2009.

\bibitem{Cooper:2012bd}
I.~K. Cooper, S.~F. King, and A.~J. Stuart, ``{A Golden $A_5$ Model of Leptons
  with a Minimal NLO Correction},'' {\em Nucl. Phys. B}, vol.~875,
  pp.~650--677, 2013.

\bibitem{deMedeirosVarzielas:2013tbq}
I.~de~Medeiros~Varzielas and L.~Lavoura, ``{Golden ratio lepton mixing and
  nonzero reactor angle with $A_5$},'' {\em J. Phys. G}, vol.~41, p.~055005,
  2014.

\bibitem{Gehrlein:2014wda}
J.~Gehrlein, J.~P. Oppermann, D.~Sch\"afer, and M.~Spinrath, ``{An SU(5)
  $\times$ A$_5$ golden ratio flavour model},'' {\em Nucl. Phys. B}, vol.~890,
  pp.~539--568, 2014.

\bibitem{di2015lepton}
A.~Di~Iura, C.~Hagedorn, and D.~Meloni, ``Lepton mixing from the interplay of
  the alternating group a5 and cp,'' {\em Journal of High Energy Physics},
  vol.~2015, no.~8, pp.~1--27, 2015.

\bibitem{PhysRevD.92.093008}
P.~Ballett, S.~Pascoli, and J.~Turner, ``Mixing angle and phase correlations
  from ${A}_{5}$ with generalized $cp$ and their prospects for discovery,''
  {\em Phys. Rev. D}, vol.~92, p.~093008, Nov 2015.

\bibitem{PhysRevD.92.116007}
J.~Turner, ``Predictions for leptonic mixing angle correlations and nontrivial
  dirac $cp$ violation from ${A}_{5}$ with generalized $cp$ symmetry,'' {\em
  Phys. Rev. D}, vol.~92, p.~116007, Dec 2015.

\bibitem{li2015lepton}
C.-C. Li and G.-J. Ding, ``Lepton mixing in $a_5$ family symmetry and
  generalized cp,'' {\em Journal of High Energy Physics}, vol.~2015, no.~5,
  pp.~1--56, 2015.

\bibitem{lopez2019lepton}
M.~L. L{\'o}pez-Ib{\'a}{\~n}ez, A.~Melis, D.~Meloni, and O.~Vives, ``Lepton
  flavor violation and neutrino masses from $a_5$ and cp in the non-universal
  mssm,'' {\em Journal of High Energy Physics}, vol.~2019, no.~6, pp.~1--34,
  2019.

\bibitem{Novichkov:2018nkm}
P.~P. Novichkov, J.~T. Penedo, S.~T. Petcov, and A.~V. Titov, ``{Modular
  A$_{5}$ symmetry for flavour model building},'' {\em JHEP}, vol.~04, p.~174,
  2019.

\bibitem{Ding:2019xna}
G.-J. Ding, S.~F. King, and X.-G. Liu, ``{Neutrino mass and mixing with $A_5$
  modular symmetry},'' {\em Phys. Rev. D}, vol.~100, no.~11, p.~115005, 2019.

\bibitem{deMedeirosVarzielas:2022ihu}
I.~de~Medeiros~Varzielas and J.~a. Louren\c{c}o, ``{Two A5 modular symmetries
  for Golden Ratio 2 mixing},'' {\em Nucl. Phys. B}, vol.~984, p.~115974, 2022.

\bibitem{Behera:2022wco}
M.~K. Behera and R.~Mohanta, ``{Linear Seesaw in A5' Modular Symmetry With
  Leptogenesis},'' {\em Front. in Phys.}, vol.~10, p.~854595, 2022.

\bibitem{Abbas:2024bbv}
M.~A. Abbas, ``{Lepton Masses and Mixing in Modular A5 Symmetry},'' {\em LHEP},
  vol.~2024, p.~545, 2024.

\bibitem{fo11}
D.~V. Forero, S.~Morisi, M.~Tortola, and J.~W.~F. Valle, ``{Lepton flavor
  violation and non-unitary lepton mixing in low-scale type-I seesaw},'' {\em
  JHEP}, vol.~09, p.~142, 2011.

\bibitem{Hernndez2023}
A.~E.~C. Hernández, K.~N. Vishnudath, and J.~W.~F. Valle, ``Linear seesaw
  mechanism from dark sector,'' {\em Journal of High Energy Physics},
  vol.~2023, Sept. 2023.

\bibitem{Batra_2023}
A.~Batra, P.~Bharadwaj, S.~Mandal, R.~Srivastava, and J.~W.~F. Valle,
  ``Phenomenology of the simplest linear seesaw mechanism,'' {\em Journal of
  High Energy Physics}, vol.~2023, July 2023.

\bibitem{ParticleDataGroup:2018ovx}
M.~Tanabashi {\em et~al.}, ``{Review of Particle Physics},'' {\em Phys. Rev.
  D}, vol.~98, no.~3, p.~030001, 2018.

\bibitem{Krastev:1988yu}
P.~I. Krastev and S.~T. Petcov, ``{Resonance Amplification and t Violation
  Effects in Three Neutrino Oscillations in the Earth},'' {\em Phys. Lett. B},
  vol.~205, pp.~84--92, 1988.

\bibitem{Jarlskog:1985ht}
C.~Jarlskog, ``{Commutator of the Quark Mass Matrices in the Standard
  Electroweak Model and a Measure of Maximal CP Nonconservation},'' {\em Phys.
  Rev. Lett.}, vol.~55, p.~1039, 1985.

\bibitem{fernandez2016global}
E.~Fernandez-Martinez, J.~Hernandez-Garcia, and J.~Lopez-Pavon, ``Global
  constraints on heavy neutrino mixing,'' {\em Journal of High Energy Physics},
  vol.~2016, no.~8, pp.~1--31, 2016.

\bibitem{Masina:2005am}
I.~Masina and C.~A. Savoy, ``{On power and complementarity of the experimental
  constraints on seesaw models},'' {\em Phys. Rev. D}, vol.~71, p.~093003,
  2005.

\bibitem{Hisano:2001qz}
J.~Hisano and K.~Tobe, ``{Neutrino masses, muon g-2, and lepton flavor
  violation in the supersymmetric seesaw model},'' {\em Phys. Lett. B},
  vol.~510, pp.~197--204, 2001.

\bibitem{Abada:2015oba}
A.~Abada, V.~De~Romeri, and A.~M. Teixeira, ``{Impact of sterile neutrinos on
  nuclear-assisted cLFV processes},'' {\em JHEP}, vol.~02, p.~083, 2016.

\bibitem{Koike:2010xr}
M.~Koike, Y.~Kuno, J.~Sato, and M.~Yamanaka, ``{A new idea to search for
  charged lepton flavor violation using a muonic atom},'' {\em Phys. Rev.
  Lett.}, vol.~105, p.~121601, 2010.

\bibitem{Ilakovac:1994kj}
A.~Ilakovac and A.~Pilaftsis, ``{Flavor violating charged lepton decays in
  seesaw-type models},'' {\em Nucl. Phys. B}, vol.~437, p.~491, 1995.

\bibitem{PhysRevLett.110.201801}
J.~Adam {\em et~al.}, ``New constraint on the existence of the
  ${\ensuremath{\mu}}^{+}\ensuremath{\rightarrow}{e}^{+}\ensuremath{\gamma}$
  decay,'' {\em Phys. Rev. Lett.}, vol.~110, p.~201801, May 2013.

\bibitem{meucci2022meg}
M.~Meucci, ``Meg ii experiment status and prospect,'' 2022.

\bibitem{Awasthi:2013we}
R.~L. Awasthi, M.~K. Parida, and S.~Patra, ``{Neutrinoless double beta decay
  and pseudo-Dirac neutrino mass predictions through inverse seesaw
  mechanism},'' 1 2013.

\bibitem{MIYAZAKI2011251}
Y.~Miyazaki {\em et~al.}, ``Search for lepton-flavor-violating $\tau$ decays
  into a lepton and a vector meson,'' {\em Physics Letters B}, vol.~699, no.~4,
  pp.~251--257, 2011.

\bibitem{Benes:2005hn}
P.~Benes, A.~Faessler, F.~Simkovic, and S.~Kovalenko, ``{Sterile neutrinos in
  neutrinoless double beta decay},'' {\em Phys. Rev. D}, vol.~71, p.~077901,
  2005.

\bibitem{Cummins:1988dr}
C.~J. Cummins and J.~Patera, ``{POLYNOMIAL ICOSAHEDRAL INVARIANTS},'' {\em J.
  Math. Phys.}, vol.~29, pp.~1736--1745, 1988.

\bibitem{Luhn:2008ey}
C.~Luhn and P.~Ramond, ``{Quintics with Finite Simple Symmetries},'' {\em J.
  Math. Phys.}, vol.~49, p.~053525, 2008.

\end{thebibliography}
\end{document}